\documentclass[fleqn,usenatbib]{mnras}

\usepackage[T1]{fontenc}
\usepackage{ae,aecompl}
\usepackage{graphicx}
\usepackage{amsmath}
\usepackage{amssymb}
\usepackage{graphicx}
\usepackage{wrapfig}
\usepackage{epstopdf}
\usepackage{rotating}
\usepackage[export]{adjustbox}
\usepackage{pdflscape}
\usepackage{caption}

\usepackage{color}
\usepackage{lscape}
\usepackage{multirow}
\usepackage[section]{placeins}
\usepackage{setspace}
\usepackage{url}

\usepackage[dvipsnames]{xcolor}

\newcommand{\be}{\begin{equation}}                                  
\newcommand{\ee}{\end{equation}}                                    

\title[ACT CMB Lensing $\times$ BOSS]{The Atacama Cosmology Telescope: A CMB lensing mass map over 2100 square degrees of sky and its cross-correlation with BOSS-CMASS galaxies}

\author[O. Darwish, M. S. Madhavacheril, B. Sherwin et. al. ]{\parbox{\textwidth}{
\Large
Omar~Darwish,$^{1}$\thanks{od261@cam.ac.uk}
Mathew~S.~Madhavacheril,$^{2,4}$\thanks{mmadhavacheril@perimeterinstitute.ca}
Blake~D.~Sherwin,$^{1,13}$\thanks{sherwin@damtp.cam.ac.uk}
Simone~Aiola,$^{10}$
Nicholas~Battaglia,$^{22}$
James~A.~Beall,$^{14}$
Daniel~T.~Becker,$^{14}$
J.~Richard~Bond,$^{23}$
Erminia~Calabrese,$^{17}$
Steve~Choi,$^{22}$
Mark~J.~Devlin,$^{29}$
Jo~Dunkley,$^{4,3}$
Rolando~D\"unner,$^{36}$
Simone~Ferraro,$^{5}$
Anna~E.~Fox,$^{14}$
Patricio~A.~Gallardo,$^{21}$
Yilun~Guan,$^{15}$
Mark~Halpern,$^{24}$
Dongwon~Han,$^{8}$
Matthew~Hasselfield,$^{10}$
J.~Colin~Hill,$^{6,10,11}$
Gene~C.~Hilton,$^{14}$
Matt~Hilton,$^{18}$
Adam~D.~Hincks,$^{23}$
Shuay-Pwu~Patty~Ho,$^{3}$
J.~Hubmayr,$^{14}$
John~P.~Hughes,$^{27}$
Brian~J.~Koopman,$^{28}$
Arthur~Kosowsky,$^{15}$
J.~Van~Lanen,$^{14}$
Thibaut~Louis,$^{16}$
Marius~Lungu,$^{3}$
Amanda~MacInnis,$^{8}$
Lo\"ic~Maurin,$^{20}$
Jeffrey~McMahon,$^{32,33,34,35}$
Kavilan~Moodley,$^{18}$
Sigurd~Naess,$^{10}$
Toshiya~Namikawa,$^{1}$
Laura~Newburgh,$^{28}$
John~P.~Nibarger,$^{14}$
Micheal~D.~Niemack,$^{21,22}$
Lyman~A.~Page,$^{3}$
Bruce~Partridge,$^{7}$
Frank~J.~Qu,$^{1}$
Naomi~Robertson,$^{12,13}$
Benjamin~Schmitt,$^{30}$
Neelima~Sehgal,$^{8}$
Crist\'{o}bal~Sif\'{o}n,$^{26}$
David~N.~Spergel,$^{10,4}$
Suzanne~Staggs,$^{3}$
Emilie~Storer,$^{19}$
Alexander~van~Engelen,$^{25}$
Edward~J.~Wollack$^{9}$}}
\pubyear{2020}
\date{}
\begin{document}
\label{firstpage}
\pagerange{\pageref{firstpage}--\pageref{lastpage}}
\maketitle

% Abstract of the paper
\vspace{-3cm}
\begin{abstract}
We construct cosmic microwave background lensing mass maps using data from the 2014 and 2015 seasons of observations with the Atacama Cosmology Telescope (ACT). These maps cover $2100$ square degrees of sky and overlap with a wide variety of optical surveys. The maps are signal dominated on large scales and have fidelity such
that their correlation with the cosmic infrared background is clearly visible by eye. We also create lensing maps with thermal Sunyaev-Zel'dovich contamination removed using a novel cleaning procedure that only slightly degrades the lensing signal-to-noise ratio. The cross-spectrum between the cleaned lensing map and the BOSS CMASS galaxy sample
is detected at $10$-$\sigma$ significance, with an amplitude of $A=1.02 \pm 0.10$ relative to the  \emph{Planck} best-fit LCDM cosmological model with fiducial linear galaxy 
bias. Our measurement lays the foundation for lensing cross-correlation science with current ACT data and beyond.
\\
\\
\\
\end{abstract}

%________________________________________________________________

% WARNING: This section was generated by a script.
% If you want to edit it 
% please edit the corresponding spreadsheet and re-run 
% the script.
% List of institutions
\begin{scriptsize}
\begin{spacing}{1.0}{
\noindent$^{1}$Center for Theoretical Cosmology, DAMTP, University of Cambridge, CB3 0WA, UK\\
$^{2}$Centre for the Universe, Perimeter Institute, Waterloo, ON N2L 2Y5, Canada\\
$^{3}$Joseph Henry Laboratories of Physics, Jadwin Hall, Princeton University, Princeton, NJ 08544, USA\\
$^{4}$Department of Astrophysical Sciences, Princeton University, 4 Ivy Lane, Princeton, NJ, USA 08544\\
$^{5}$Physics Division, Lawrence Berkeley National Laboratory, Berkeley, CA 94720, USA\\
$^{6}$Department of Physics, Columbia University, 550 West 120th Street, New York, NY, USA 10027\\
$^{7}$Department of Physics and Astronomy, Haverford College,Haverford, PA, USA 19041\\
$^{8}$Physics and Astronomy Department, Stony Brook University, Stony Brook, NY 11794, USA\\
$^{9}$NASA/Goddard Space Flight Center, Greenbelt, MD 20771, USA\\
$^{10}$Center for Computational Astrophysics, Flatiron Institute, 162 5th Avenue, New York, NY, USA 10010\\
$^{11}$School of Natural Sciences, Institute for Advanced Study, Princeton, NJ, USA 08540\\
$^{12}$Institute of Astronomy, Madingley Road, Cambridge CB3 0HA, UK\\
$^{13}$Kavli Institute for Cosmology, University of Cambridge, Madingley Road, Cambridge CB3 OHA, UK\\
$^{14}$NIST Quantum Devices Group, 325 Broadway Mailcode 817.03, Boulder, CO, USA 80305\\
$^{15}$Department of Physics and Astronomy, University of Pittsburgh, Pittsburgh, PA, USA 15260\\
$^{16}$Universit\'e Paris-Saclay, CNRS/IN2P3, IJCLab, 91405 Orsay, France\\
$^{17}$School of Physics and Astronomy, Cardiff University, The Parade, Cardiff, CF24 3AA, UK\\
$^{18}$Astrophysics \& Cosmology Research Unit, School of Mathematics, Statistics \& Computer Science, University of KwaZulu-Natal, Westville Campus, Durban 4041, South Africa\\
$^{19}$Joseph Henry Laboratories of Physics, Jadwin Hall, Princeton University, Princeton, NJ 08544, USA\\
$^{20}$Universit\'e Paris-Saclay, CNRS, Institut d'astrophysique spatiale, 91405, Orsay, France\\
$^{21}$Department of Physics, Cornell University, Ithaca, NY 14853 USA\\
$^{22}$Department of Astronomy, Cornell University, Ithaca, NY 14853 USA\\
$^{23}$Canadian Institute for Theoretical Astrophysics, University of Toronto, 60 St. George Street, Toronto, ON, Canada, M5S 3H8\\
$^{24}$Department of Physics and Astronomy, University of British Columbia, Vancouver, BC V6T 1Z1, Canada
\\
$^{25}$School of Earth and Space Exploration and Department of Physics, Arizona State University, Tempe, AZ 85287\\
$^{26}$Instituto de F\'isica, Pontificia Universidad Cat\'olica de Valpara\'iso, Casilla 4059, Valpara\'iso, Chile\\
$^{27}$Department of Physics and Astronomy, Rutgers University, 136 Frelinghuysen Road, Piscataway, NJ 08854-8019 USA\\
$^{28}$Department of Physics, Yale University, New Haven, CT 06520, USA\\
$^{29}$Department of Physics and Astronomy, University of Pennsylvania, 209 South 33rd Street, Philadelphia, PA 19104\\
$^{30}$Harvard-Smithsonian Center for Astrophysics, Harvard University, 60 Garden St, Cambridge, MA 02138, United States\\
$^{31}$The University of Michigan Department of Physics 450 Church Street, Ann Arbor, Michigan, 48109\\
$^{32}$Kavli Institute for Cosmological Physics, University of Chicago, 5640 S. Ellis Ave., Chicago, IL 60637, USA\\
$^{33}$Department of Astronomy and Astrophysics, University of Chicago, 5640 S. Ellis Ave., Chicago, IL 60637, USA\\
$^{34}$Department of Physics, University of Chicago, Chicago, IL 60637, USA\\
$^{35}$Enrico Fermi Institute, University of Chicago, Chicago, IL 60637, USA\\
$^{36}$Instituto de Astrof\'isica and Centro de Astro-Ingenier\'ia, Facultad de F\'isica, Pontificia Universidad Cat\'olica de Chile, Av.  Vicu\~na Mackenna 4860, 7820436 Macul, Santiago, Chile\\

}
\end{spacing}
\end{scriptsize}
% END OF WARNING

\section{\label{sec:level1}Introduction}

Along their paths to our telescopes, the photons of the cosmic microwave background (CMB) are deflected, or lensed, by the gravitational influence of the matter in our Universe. This leads to a remapping of the observed CMB anisotropies on the sky described by $T(\mathbf{\mathbf{\hat{n}}})=T^u(\mathbf{\mathbf{\hat{n}}}+\mathbf{d})$, where $T$ and $T^u$ are the lensed and unlensed temperature fields and $\mathbf{\mathbf{\hat{n}}}$ is the line of sight. (Analogous expressions hold for the remapping of polarization $Q$ and $U$). The lensing deflection field $\mathbf{d}(\mathbf{\mathbf{\hat{n}}})$ that describes the remapping depends on a weighted integral of the mass along the line of sight; although this integral extends to the last-scattering surface, most of the lensing signal arises between redshifts $z=0.5$ and $z=3$ \citep[][]{astro-ph/9810257, LEWIS2006}. Since maps of the CMB lensing signal are sensitive to the total matter distribution, including dark matter, they contain a wealth of information about cosmology and fundamental physics \citep[e.g.,][]{Lesgourgues2006, Sherwin2011, PlanckLensing:2018}.

In this paper, we present a CMB lensing map constructed from new observations from ACT, which will be useful for cross-correlation analyses.

Cross-correlation measurements can be used to break the degeneracy of galaxy bias (the factor relating the galaxy and matter density contrasts) and
the amplitude of matter density fluctuations. This allows us
to determine the amplitude of structure at different redshifts $\sigma_8(z)$ \citep[e.g.,][]{Giannantonio, Peacock, Giusarma:2018, Doux} and hence probe physics such
as dark energy, modified gravity, and neutrino mass. CMB lensing cross-correlations can also be used to
constrain multiplicative biases in shear measurements \citep[e.g.,][]{1110.5339,1311.2338,HandXCorr,1601.05720,1607.01761}, measure cosmographic
distance ratios \citep[e.g.,][]{0708.4391,0810.3931,Miyatake:2017,1810.02212},
calibrate the masses of galaxy groups and clusters \citep[e.g.,][]{1411.7999,1412.7521, 1408.5633,1502.01597,1904.07887,1907.08605,1708.01360,1907.08605,Raghunathan2019},
and probe astrophysics via the relation of dark to luminous matter \citep[e.g.,][]{Sherwin:2012,bleem2012,1412.0626,1502.06456,1303.5078,1902.06955,1809.04196,1711.10774,Omori2017,1710.09770}. However, a key challenge in such analyses is that CMB lensing maps reconstructed from temperature anisotropies can be contaminated by foreground emission and scattering \citep{SmithForegrounds,HirataForegrounds,vanEngelenForegrounds,DasForegrounds,kSZBias}, which can induce $10-20\%$ level biases in the measured cross-correlation signal \citep{OmoriForegrounds,Baxter2018}. For cross-correlations with low-redshift tracers, these foreground biases arise predominantly from the thermal Sunyaev-Zel'dovich (tSZ) residuals that lie in the map.

To solve this problem, in this paper we develop and implement a new cleaning method, building on \cite{Madhavacheril:2018} (hereafter MH18), in order to eliminate foregrounds from the tSZ effect in cross-correlations. The foreground removal in our method is achieved while preserving nearly all of the cross-correlation signal-to-noise.

We demonstrate the potential of our new foreground-cleaned CMB lensing maps, which overlap with a variety of optical surveys, by measuring a robust cross-correlation of these maps with Sloan Digital Sky Survey DR12 BOSS CMASS spectroscopic galaxies \citep[][]{BOSSDR12}.

We also note that some analyses found a lower cross-correlation spectrum between CMB lensing and both low redshift galaxies and weak lensing than expected from the \emph{Planck} cosmology \citep[e.g., ][]{Pullen:2016,Liu}. Testing this possible discrepancy with our new lensing maps provides further motivation for our analysis.

Our paper is structured as follows. Section 2 explains the theoretical background for our cross-correlation measurement. In Section 3 we present our data and discuss the new lensing maps constructed from ACT data. In Section 4 we discuss the construction of tSZ-free lensing maps.  In Section 5 we present the cross-correlation measurement with CMASS BOSS galaxies, followed by a discussion of systematic errors in Section 6. The conclusions follow in the final section of our paper. Two appendices explain the CMB map pre-processing and discuss, in more detail, the cleaning method used to remove the tSZ bias from the lensing maps.

%__________________________________________________________________

\section{\label{sec:Theory}Theoretical Background}
The CMB lensing convergence field $\kappa$, which is related to the lensing deflection via $\kappa = \frac{1}{2} \nabla \cdot \mathbf{d}$, is a direct measure of the projected matter field. In particular, the convergence can be shown to equal a weighted integral of the matter density perturbation along a line of sight with direction $\mathbf{\hat{n}}$:
\begin{equation}
    \kappa(\mathbf{\hat{n}})=\int_0^{z_*}dzW^{\kappa}(z)\delta(\chi(z)\mathbf{\hat{n}}, z)
\end{equation}
with $z_*$ the redshift at the last scattering surface, $\delta$ the three
dimensional matter density contrast field at redshift $z$, $\chi(z)$ the comoving distance at redshift $z$, and the window response kernel $W^{\kappa}$ for redshift $z$ given by \citep[e.g.,][]{Sherwin:2012}
\begin{equation}
    W^{\kappa}(z)=\frac{3}{2H(z)}\Omega_{m,0}H_0^2(1+z)\chi(z)\frac{\chi_*-\chi(z)}{\chi_*},
\end{equation}
where $H(z)$ is the Hubble parameter as a function of redshift, $H_0$ its value today, $\chi_*=\chi(z_*)$,and $\Omega_{m,0}$ is the value of the matter density parameter today. 

The 3D distribution of galaxies can provide an independent view of the matter distribution in combination with lensing, and one that can probe the time dependence of structure growth. (In contrast, $\kappa$ is a projection of the matter field over a very wide range of redshifts and so cannot provide tomographic information.) The relevant cosmological field is the fractional number overdensity of galaxies in a direction $\mathbf{\hat{n}}$, given by another weighted integral along the line of sight
\begin{equation}
    \delta_g(\mathbf{\hat{n}})=\int_{0}^{z_*} dz W^g(z)\delta^{ \mathrm{3D}}_g(\chi(z)\mathbf{\hat{n}}, z),
\end{equation} 

\noindent where $\delta^{ \mathrm{3D}}_g$ is
the three dimensional galaxy distribution at redshift $z$ and the window function $W^g(z)$ is $\frac{dn}{dz}(z)$, the
redshift distribution of galaxies in a galaxy survey, normalized to unity.\footnote{We do not
  include magnification bias, since its magnitude is negligible
  given the low redshift range of the galaxy catalog used in this work.} In this work, we consider a spectroscopic galaxy survey with a redshift-binned sample such that the kernel $W$ is only non-zero between $z_i$ and $z_f$, with $z_i$, $z_f$ the low and high redshifts defining the survey.

Since galaxies are biased tracers of the underlying matter distribution, the matter-galaxy power spectrum is
\begin{equation}
P_{mg}(k,z) = b_{\rm cross}(k,z)P(k,z)
\end{equation}
where $b_{\rm cross}(k,z)$ is a general scale- and redshift-dependent clustering bias and $P(k,z)$ is the matter power spectrum \citep[][]{Blanton_1999}. In our cross-correlation analysis, we explicitly choose the scales and redshift-range included such that the scale- and redshift-dependence of the galaxy bias is not large and $b_{\rm cross}(k,z)\approx b= \rm const$. We will consider multipoles $L$ in the range $100<L<1000$; this choice will be motivated in Section 5.

The cross-power spectrum of the two observables $\kappa$ and $g$ is directly related to the cosmological parameters of the underlying $\Lambda$CDM model. Using the flat-sky approximation valid for a  small sky
fraction $f_{\rm sky}$ and the Limber approximation \citep{Limber:1953}, the expression for the cross-spectrum in the linear
$\Lambda$CDM model is \citep[e.g.,][]{OmoriHolder}:

\begin{equation}
C_L^{\kappa g} = \int_0^{z_{*}} dz \frac{H(z)}{\chi^2(z)} W^\kappa(z) \frac{dn}{dz}(z) P_{mg}\left(k=\frac{L+\frac{1}{2}}{\chi(z)}, z\right)
\label{eq:clkg}\ .
\end{equation}

%__________________________________________________________________

\section{\label{sec:Data}Lensing Maps from ACT Data Alone}

We construct two CMB lensing maps. The first map, described in this section, uses ACT data alone. The second, described in the following section, also uses multi-frequency data from \emph{Planck} in order to clean foregrounds.

\subsection{CMB maps for lensing analysis}

The lensing convergence maps used in this work are constructed from CMB
temperature and polarization data taken by the polarization-sensitive receiver on
the Atacama Cosmology Telescope (ACT), a 6-meter CMB telescope operating
in the Atacama desert in Chile \citep[see e.g., ][]{Thornton, Choi:2019, Aiola:2019}. The CMB field maps are obtained from observations made
during seasons $2014-2015$ in the $98$ GHz and $150$ GHz frequency bands; these maps will be made public, along with our lensing maps, in the upcoming ACT data release 4 (DR4). We will consider data coming from two regions of the sky, one referred to as BOSS-North or {\tt BN} (from the 2015 season, covering $\approx 1633$ sq. deg. of the sky
overlapping the SDSS BOSS northern field, with effective co-added white noise
level of approximately $\Delta_T=21\mu $K-arcmin for temperature and
$\Delta_P=\sqrt{2}\Delta_T$ for polarization), and the other referred to as  {\tt D56}
(seasons 2014-2015, covering $\approx 456$ sq. deg. of the sky, with effective co-added white noise level of approximately $\Delta_T=10\mu $K-arcmin for temperature and $\Delta_P=\sqrt{2}\Delta_T$ for polarization).\footnote{Atmospheric noise contributes a $1/f$ component that is non-negligible and must be included when forecasting the signal-to-noise in the lensing map.} Given the proximity of the maps to the equator and their moderate extent in declination, the flat-sky approximation is sufficient at our accuracy for constructing lensing maps; a simple estimate of the inaccuracy of this approximation gives no detectable effect for {\tt D56} and only a 1\% multiplicative bias for {\tt BN}. We do not use 2013 or 2016 observations in our analysis (even though the latter are part of DR4), because the 2013 observations cover too little sky area and the 2016 observations are still too shallow to contribute significant signal-to-noise to cross-correlation measurements.

We combine the per-season and per-frequency CMB maps presented in \cite{Choi:2019} to provide the input maps for our lensing estimator. The details of this procedure are described in Appendix A, but we briefly summarize them here. We construct our CMB input maps by co-adding source-subtracted\footnote{See \cite{Madhavacheril:2019}, \cite{Choi:2019}, \cite{Aiola:2019} for details.} maps from the two frequencies and two seasons of the data and convolving the result to a common beam after masking. In addition, we inpaint (fill with an appropriately correlated Gaussian random field) a 6-arcmin-radius circular area around bright compact sources and SZ clusters using the maximum likelihood method of \cite{Bucher:2012}. This inpainting step serves to reduce foreground biases arising from bright sources and massive clusters. We note that the main difference from the map processing employed in \cite{Sherwin:2017} is that the different frequencies and seasons are coadded with weights that are local in Fourier space rather than real space; this is more optimal for multifrequency data due to the strong frequency dependence of the beams.

\def\ellb{{\boldsymbol{ \ell}}}                                                      
\def\Lb{{\boldsymbol{ L}}}

The results of our map construction and preparation process are masked, beam-deconvolved dimensionless CMB fluctuation
      maps of temperature $T$ as well as $Q$ and $U$ polarization in each of the two sky regions. The $Q$ and
      $U$ polarization maps are transformed into $E-B$ polarization maps using the pure $E-B$
      decomposition method outlined in \cite{Louis:2013}.
      As a final step in the preparation of the maps for lensing reconstruction, we follow the nominal analysis methodology of \cite{Choi:2019} to reduce the impact of
      ground contamination in the $T$, $E$ and $B$ maps, filtering out all modes $\ellb=(\ell_x,
    \ell_y)$ that have $|\ell_x|<90$ and $|\ell_y|<50$. We also remove all
    modes that are outside the range of scales $500<\ell<3000$ in order to restrict our lensing analysis to scales
    where the ACT map-maker transfer function is small\footnote{The map-maker transfer function is close to unity for $\ell>500$ in {\tt D56}, but may be as large as 10\% in {\tt BN} between $\ell$ of 500 and 600 \citep[]{Choi:2019,Aiola:2019}. However, because of the fact that the lensing estimator only draws a small fraction of its statistical weight from multipoles $500 < \ell<600$ (less than 2\%, see e.g. \cite{Schmittfull2013}), we expect an effect on lensing cross-correlations that is much smaller than the statistical uncertainty and is thus negligible.} and 
    where contamination from foregrounds is small ($\ell<3000$).
    
    As well as processing data, we also produce $N=511$ CMB simulations matching each of the CMB maps described above. These simulations
are generated using the pipeline described in \cite{Choi:2019} and include
primary CMB, lensing, noise and foregrounds. The foregrounds are Gaussian and spatially homogeneous and the noise is Gaussian but spatially inhomogeneous, as described in \cite{Choi:2019}. We use the simulations to test our lensing reconstructions, derive small transfer function corrections and
construct covariance matrices, as described in the following sections of this paper. To reconstruct lensing convergence maps from simulations we use the same pipeline that we apply to the data. We describe this lensing reconstruction pipeline in the following subsection.

    \subsection{Lensing reconstruction and validation}

    Exploiting the mode couplings induced by lensing, we reconstruct the lensing convergence field from our CMB maps with a minimum variance quadratic estimator \citep[][]{Hu:2002}:

\begin{equation}
    \bar{\kappa}^{XY}(\Lb) = A^{XY}(\Lb)\int \frac{d^2\ellb}{(2\pi)^2}X(\ellb) Y(\Lb-\ellb)f^{XY}(\ellb, \Lb)    \label{eq:kappaquadratic}
\end{equation}
where $A^{XY}(\Lb)$ is a normalization (derived from our fiducial cosmology) to ensure that the estimator is unbiased. $f^{XY}(\ellb, \Lb)  $ is an optimal weighting function chosen to minimise the reconstruction noise of the estimator; it includes a Wiener filter for the CMB input fields $X, Y$. As in \cite{Sherwin:2017} we will consider only the pairs $XY \in \{TT, TE, EE, EB\}$, as the $TB$ combination has negligible signal-to-noise. Expressions for the weighting function $f$ and the theory normalization $A$ can be found in \cite{Hu:2002}, although following \cite{Hanson:2011} we replace the unlensed spectra with lensed spectra in the weighting functions to cancel higher-order biases. A spurious signal on the largest scales of the reconstructed lensing map arises from non-lensing statistical anisotropy due to sky masks or inhomogeneous map noise; this spurious lensing ``mean field'' must be subtracted from Equation \ref{eq:kappaquadratic} \citep[e.g.,][]{Namikawa:2013}. We calculate this mean field correction by generating 511 lensing reconstructions from simulations and averaging these reconstructions. We thus obtain the mean-field subtracted lensing convergence estimator
\begin{equation}
    \hat{\kappa}^{XY}(\Lb)=\bar{\kappa}^{XY}(\Lb)-\langle {\kappa_s^{XY}}(\Lb) \rangle_s,
\end{equation}{} 
\noindent where ${\kappa_s^{XY}}(\Lb)$ is the lensing reconstruction $\bar{\kappa}^{XY}$ for the
simulation realization $s$ and the angle average $\langle  \rangle_s$ is over simulations.

We complete the lensing map by creating a minimum variance combination of the different types of quadratic estimators $XY \in \{TT, TE, EE, EB\}$,
\begin{equation}
    \hat\kappa^{\rm MV}_{\Lb}=\sum_{X Y}w^{XY}(\Lb)\hat\kappa^{XY}(\Lb),
    \label{eq:mvlensingnomc}
\end{equation}
where $w^{XY}(\Lb)$ are minimum variance weights.

\begin{figure}
    \centering
    \includegraphics[width=9cm]{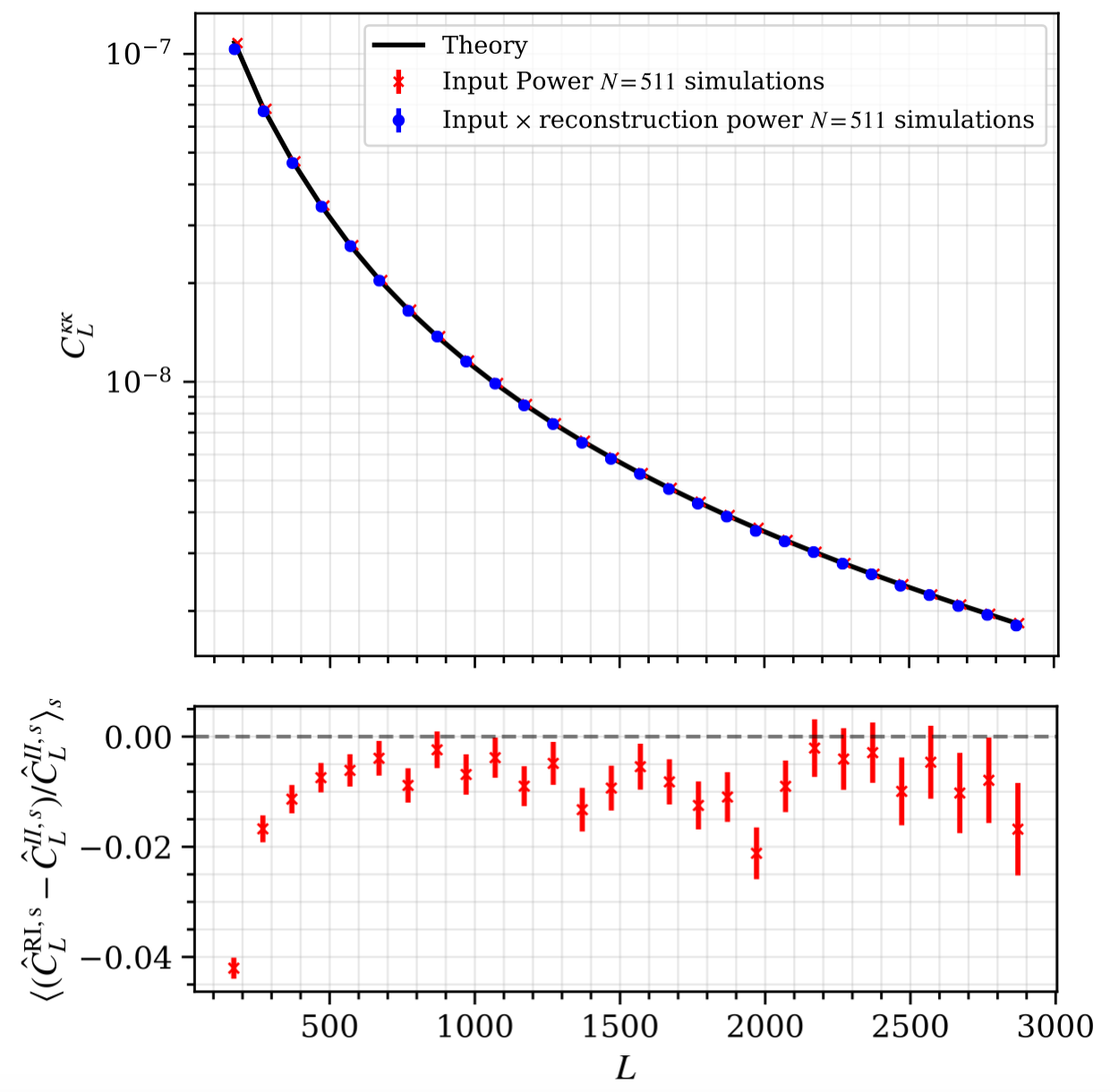}
    \caption{Verification of our lensing reconstruction pipeline for the tSZ free lensing maps (shown for the {\tt D56} patch). We plot the average cross-spectrum of the reconstructed lensing maps with the input lensing simulations (blue dots), the average power spectrum of the input lensing simulations (red crosses) and a binned lensing power theory curve in black. (The {\tt BN} patch gives quantitatively similar results.) The bottom panel shows the fractional difference of the input-reconstruction cross-correlation relative to the input lensing power. The ACT only simulations give residuals of similar magnitude. From the good agreement of the input-reconstruction cross-correlation with the input lensing power, we can see that the true lensing signal in the simulations is recovered within percent-level accuracy; we absorb only a small correction into a simulation-based re-normalization.}
    \label{fig:lensverification}
\end{figure}{}

Finally, the particular form of the normalization $A^{XY}(\Lb)$ used in
Equation \ref{eq:kappaquadratic} is valid for CMB maps with
periodic boundaries. This is clearly an idealization; for example, using
masked CMB maps introduces spurious gradients at the mask
boundary~\citep{Hirata:2008}, changing the form of the correct lensing
normalization (although this effect is reduced by apodization). We capture this
and other non-idealities by introducing an extra multiplicative normalization
function $r^{\rm MC}(L)$. 

\begin{landscape}
\pagestyle{plain}
\begin{figure*}
    \vspace*{-0.3cm}
    \includegraphics[width=1.1\linewidth,trim=14.5cm 0 0 -3cm]{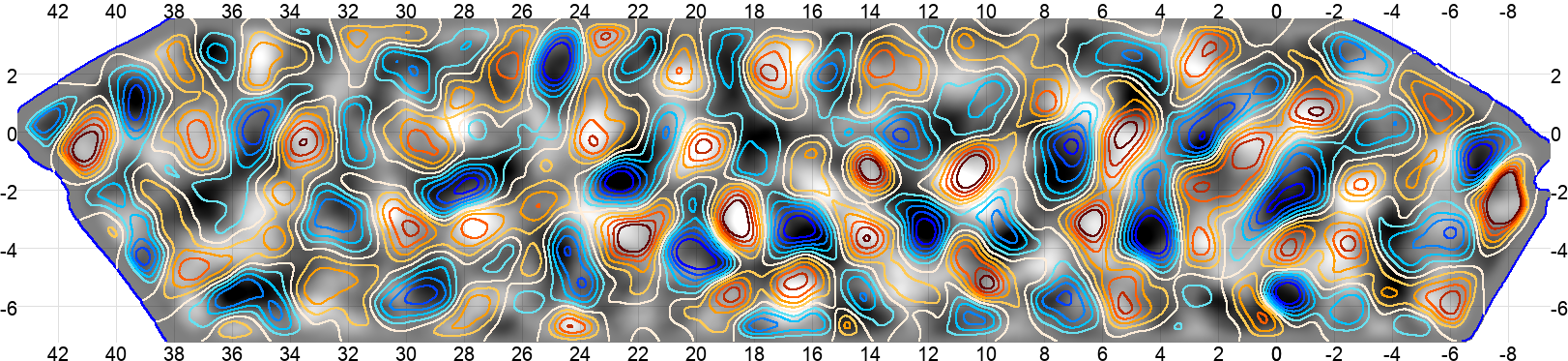} \\
    \vspace*{0.7cm}
    
     \hspace*{-4cm}
    \sbox0{{{\includegraphics[width=1.49\linewidth,trim=0cm 0 0 0]{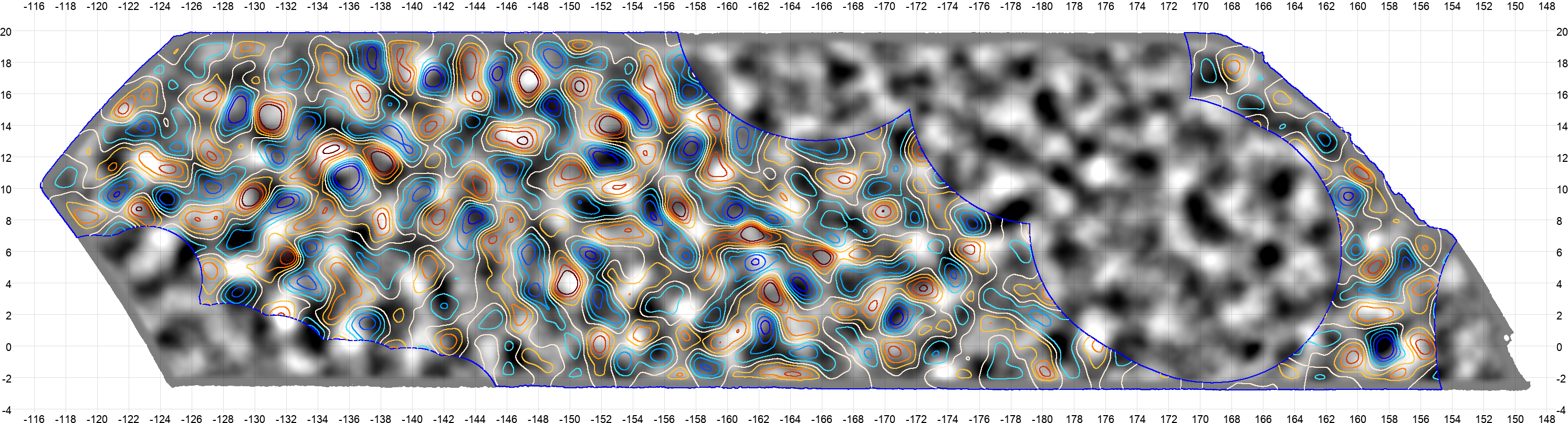}} }}
    
\hspace*{-6.8cm}\begin{minipage}{\wd0}
  \usebox0
  \caption{Map of the reconstructed lensing potential in the {\tt D56} region
    (upper panel) and the {\tt BN} region (lower panel) after Wiener
    filtering, shown in greyscale. (The lensing maps shown are the tSZ-cleaned maps combining \emph{Planck} and ACT, although the ACT-only lensing maps appear similar.) Overlaid, we also show contours of an
    identically filtered but completely independent cosmic infrared background map (\emph{Planck} GNILC 545 GHz). Since the
    correlation between CMB lensing and the cosmic infrared background (CIB) is very high and since our CMB
    lensing map has high signal-to-noise ratio on large scales, the
    correspondence between the lensing potential and the CIB can be seen
    clearly. Parts of the CIB map contaminated by Galactic dust have been masked in the {\tt BN}
    CIB contours for this visualization, using a mask derived from the \emph{Planck} PR2 Commander thermal dust emission map.}
    \label{fig:lmap}
\end{minipage}
    
\end{figure*}
\end{landscape}

To calculate this function, we cross-correlate our $N=511$ reconstructed lensing simulations $\hat{\kappa}^{\rm MV}_s$ with the true input lensing convergence field  $\kappa_s$
used to generate the simulations,\footnote{To mimic the processing of the reconstructions we mask
  $\kappa_s$ with the square of the data-mask, as this enters twice in the
  quadratic lensing estimator used to reconstruct the lensing simulation.}
obtaining the reconstruction-input cross-spectrum $\hat{C}_{L}^{\rm RI,s}$. We compare this cross-spectrum
with  the auto-spectrum of the input convergence field $\hat{C}_{L}^{\rm II,s}$. Taking the ratio of averages over the $N$
sims $\langle \hat{C}_{L}^{II,s}\rangle_s/\langle\hat{C}_{L}^{\rm RI,s}\rangle_s$,
we obtain a one dimensional binned function of $L=|\Lb|$, where $L_{\rm
  min}=20$, 
$L_{\rm max}=3000$, and 
$\Delta L = 100$. We then interpolate this over a two dimensional grid
to get the final isotropic correction function $r^{\rm MC}(\Lb)$ that we apply to the lensing maps to obtain the MC corrected minimum variance lensing maps

\begin{equation}
    \hat\kappa_{\Lb}=r^{\rm MC}(\Lb)\hat\kappa^{\rm MV}_{\Lb}
    \label{eq:mvlensing}\ .
\end{equation}{}

If our pipeline is estimating the lensing signal reliably, the Monte-Carlo based normalization
correction of Equation \ref{eq:mvlensing} should only require a rescaling of
order a few percent. To validate our pipeline, we therefore test whether our lensing map is nearly
correctly reconstructed even in the absence of Monte-Carlo renormalization.

In
Figure \ref{fig:lensverification} we show a comparison between $\langle
C_{LL}^{\rm RI,s} \rangle_s$ and $\langle C_{L}^{\rm II,s} \rangle_s$ for the {\tt D56} patch without the Monte-Carlo normalization (this figure uses foreground-cleaned ACT+\emph{Planck} lensing maps that we will introduce in the next section, but the residuals for the ACT-only maps are similar). We recover the signal with only percent-level deviations (which implies that $r^{\rm MC}(\Lb)$ is within a few percent of unity); this gives confidence that our pipeline is functioning correctly. We obtain quantitatively similar results for the {\tt BN} patch.

\subsection{Visualization of the maps and their correlation with large-scale structure}
An image of the ACTPol CMB lensing maps is shown in Figure \ref{fig:lmap}. The maps have been Wiener filtered to show the signal-dominated scales (roughly 1 degree or larger for {\tt BN} and 0.5 degrees or larger for {\tt D56}) and have been converted to maps of the lensing potential using the appropriate filtering. We also overplot contours of Cosmic Infrared Background (CIB) emission obtained from the GNILC \emph{Planck} component separated maps \citep{GNILC}; the CIB maps have the same filtering applied as the lensing ones. In the {\tt BN} region, we mask the CIB map using the \emph{Planck} PR2 Commander high-resolution map of thermal dust emission \citep[]{PlanckDust}. The mask is made by thresholding the dust map such that it covers regions of the CIB map that have visibly low power due to dust contamination; we only use this mask for the visualization of Figure \ref{fig:lmap}. The CIB arises from similar redshifts as CMB lensing and hence is known to be highly correlated with lensing \citep{Song_2003, Gil, Planckcib2013}. Indeed, even by eye a high correlation of our lensing maps and the CIB is visible. This illustrates the fact that our lensing maps are signal-dominated over a range of large scales and are a faithful tracer of the mass distribution. 

%______________________________________________________________

\section{\label{sec:fgmitigation} Foreground-mitigated Lensing maps with New Cleaning Methods}

CMB temperature maps contain secondary anisotropies not only from lensing, but also from tSZ, CIB (Cosmic Infrared Background), kSZ (kinetic Sunyaev-Zel'dovich), and other foreground contributions arising from a wide range of redshifts. The lensing estimator is sensitive to these extragalactic foregrounds \citep[see][]{vanEngelenForegrounds,Osborne,kSZBias}, which can be problematic: foreground contamination which has leaked through the lensing estimator can correlate with the galaxy distribution, giving spurious biases to cross-correlation measurements. It is important to mitigate these foregrounds in temperature, as many current- and next-generation lensing maps will still depend to a large extent on temperature data, rather than on polarization. Indeed, for our current dataset, the temperature ($TT$) lensing estimator still provides the dominant contribution ($>50\%$) to our minimum variance lensing estimate of Equation \ref{eq:mvlensingnomc}.

One of the primary goals of making a lensing map is to enable cross-correlation science. For low-$z$ large-scale-structure tracers,
such as the CMASS galaxies used in later sections of this paper, the main
contribution to the cross-correlation bias comes from the tSZ contamination of
the temperature maps
\citep{vanEngelenForegrounds,Baxter2018,Madhavacheril:2018}. The tSZ is most important because, while the tSZ and the CIB can both be significant contaminants, the CIB only weakly correlates with low-$z$ galaxies (as only a small fraction of the CIB arises from low redshifts).

The observed,
SZ contaminated temperature map, denoted  $T_{\rm{with-sz}}$, now includes an SZ
contribution $T_{\rm{tSZ}}$, so that $T_{\rm{with-sz}} = T_{\rm{cmb}}+T_{\rm{tSZ}}$.\footnote{The observed temperature map clearly also has other contributions in addition to $T_{\rm{cmb}}$ and $T_{\rm{tSZ}}$, but our focus here will be just on these two components.} When inserting this CMB map into
a quadratic lensing estimator $\hat\kappa(T_{\rm{with-SZ}}, T_{\rm{with-SZ}})$ and
cross-correlating the resulting lensing map with a galaxy map $g$, the
cross-correlation is now biased by a new bispectrum term of the form $\langle gT_{\rm{tSZ}}T_{\rm{tSZ}}
\rangle$.

For typical cross-correlations, this effect can be significant,
giving biases up to a $10-20\%$ level on large scales \citep{OmoriForegrounds,Baxter2018}; the sign of the effect is
typically negative on large scales, so that a cross-correlation with a tSZ-contaminated lensing map is biased low.\footnote{A physical explanation for this
  negative bias effect is the following. Consider a direction in which there is
  a long wavelength overdensity. Due to non-linear evolution and mode coupling,
  small-scale tSZ fluctuations are also enhanced in this direction, which
  increases the CMB temperature power at small scales, $l>2000$. This excess
  small-scale power is similar in effect to an overall `shift' of the primary
  CMB towards smaller scales. The lensing estimator interprets this locally as
  arising from demagnification due to a matter underdensity: cross-correlating
  this spurious underdensity lensing signal with the distribution of galaxies
  (which trace the overdensity) therefore results in a negative
  cross-correlation\citep{vanEngelenForegrounds}.}

Since low cross-correlations were found in several analyses
\citep[e.g.,][]{Pullen:2016}; it is interesting to consider if this type of contamination
could have an impact on previously published cross-correlation
measurements. However, we note that most of the analyses with low
cross-correlations used \emph{Planck} lensing maps. For \emph{Planck}, such
foreground biases are expected to be much less problematic (due to the lower
experimental angular resolution).

\subsection{A new tSZ-free estimator}

To account for the potential problem of tSZ contamination, we attempted to use the method of MH18 to remove foreground contamination. However, this method did not perform as well as expected. We therefore developed a new foreground-cleaned lensing estimator, extending and revising the MH18 method; we will explain the relevant details in the following paragraphs.

The basic goal of our foreground-cleaning approach is to remove foreground contamination without assuming a model for
the foregrounds' statistical properties, relying instead on the fact that the
foregrounds' frequency dependence differs from that of the CMB. A simplistic frequency cleaning of the CMB maps, however, typically degrades the lensing signal-to-noise. MH18 uses the standard lensing convergence quadratic estimator written in
real space in a form where a gradient and a non-gradient field can be distinguished \citep[e.g.,][]{Hu:2007,
LEWIS2006}. Usually, for the temperature quadratic estimator $\hat \kappa(T_1, T_2)$, the two
fields $T_1, T_2$ are chosen to be identical. However, one may, of course, use two
different CMB temperature maps in the estimator; the two maps could be processed differently or even come from
different surveys. In particular, since the spectral energy distribution (SED) of the tSZ
effect is known to high accuracy (barring relativistic and multiple-scattering effects), CMB maps made from multi-frequency data that explicitly
null or deproject the tSZ can be made. Such maps generally have higher
noise. In the procedure suggested by MH18, it is pointed out that even if only one of the two fields in the quadratic estimator is free from
tSZ, then the resulting lensing map cross-correlation will still have zero
tSZ contamination, while the noise increase due to foreground cleaning will only be moderate (since only one noisy cleaned map is used, instead of two). One way of understanding this is to note that, since the
cross-correlation bias arises from a foreground-foreground-galaxy bispectrum $\langle g T_{\rm{tSZ}}T_{\rm{tSZ}}  \rangle$,
nulling even one of the foreground fields sets the whole bispectrum $\langle gT_{\rm{tSZ}} 0  \rangle$ to zero, which gives an effectively bias-free cross-correlation measurement. We denote this foreground-cleaned MH18 estimator as $\hat\kappa(T_{\rm{no-tSZ}}, T_{\rm{with-tSZ}})$ (where the first map is
the gradient field in the lensing estimator). Despite the use of a
noisier tSZ-deprojected map in one field of the quadratic estimator, the loss in signal-to-noise in constructing this foreground-free lensing map was claimed in MH18 to be only $\approx 5\%$.

However, when implementing the MH18 estimator, we found that
the actual lensing map noise obtained in both simulations and in data was larger for $L<800$ (by more than an order of magnitude at $L\approx 100$, see Figure \ref{fig:noiseforecastsTT}) than the noise forecast
    presented in MH18. The explanation for this result is the following: in MH18 a simplified formula for the noise forecast was used (namely assuming the noise is equal to the normalization, i.e. $N_L\propto L^2 A_L$); however, this is only valid if the weights in the estimator are minimum-variance. As detailed in Appendix B, the MH18 estimator does not use minimum-variance weights, which explains why the true noise we find is larger than the simplified forecast results. We note that the MH18 forecast is however accurate for cluster scales, where the gradient approximation holds in the squeezed limit \citep[]{Hu:2007,Raghunathan2019}.

To solve the problem of increased noise on large scales, we propose a new `symmetrized' cleaned estimator, in which
we coadd the $\hat\kappa(T_{\rm no-tSZ},
T_{\rm with-tSZ})$ MH18 estimator with a version where the two fields have been permuted, $\hat\kappa(T_{\rm with-tSZ}, T_{\rm no-tSZ})$. In particular, we define $\hat\kappa^{TT}_{\rm symm,~tSZfree} = \sum w_\alpha(\Lb)\hat{\kappa}_{\alpha}(\Lb)$ with weights

\begin{equation}
\label{eq:weights}
    w_{\alpha}(\Lb)=\frac{\sum_{\beta}N^{-1}_{\alpha\beta}(\Lb)}{\sum_{\gamma, \beta}N^{-1}_{\gamma\beta}(\Lb)}
\end{equation}{}
where $\alpha \in \{(T_{\rm no-tSZ},T_{\rm with-tSZ}), (T_{\rm with-tSZ},T_{\rm no-tSZ})\}$ and $N^{-1}$ is the inverse $2 \times 2$ covariance matrix taking into account the cross-correlation between the two estimators.

The resulting $\hat\kappa^{TT}_{\rm symm,~tSZ-free}$ map retains the property
that the resulting cross-correlation with large-scale structure is
unbiased, but the lensing map now has significantly lower noise: in fact, we find that our method appears to effectively recover the
original forecast results of MH18, primarily due to the
cancellation of anti-correlated noise on large scales from each of the two terms in the new estimator. Details can be found in Appendix B.

\subsection{Application to data}

The above technique requires maps of the CMB in which the tSZ signal has been
deprojected (i.e., nulled) using multi-frequency data. Such maps were presented in \cite{Madhavacheril:2019}; these maps were constructed by combining
\emph{Planck} and ACT\footnote{Despite including \emph{Planck} data, in these
  maps, the
  small-scales relevant for lensing are dominated by the ACT 148 GHz and 97 GHz channels.} data using an internal linear combination
(ILC) algorithm. We use the constrained ILC CMB map (with tSZ deprojection) and
the standard ILC CMB map (with no deprojection)\footnote{We use version v1.1.1 of the maps for which bandpass corrections for the tSZ response may not be accurate at the few percent level at the map-level. However, since the tSZ bias is at most 20\% in power, tSZ-cleaned cross-correlations are only affected at the 1\% level, an order of magnitude below the statistical sensitivity of this work.} from that analysis as the two input maps for the symmetrized cleaned
lensing estimator $\hat\kappa^{TT}_{\rm symm,~tSZ-free}$ described above; we thus create new foreground-cleaned temperature lensing maps.\footnote{Before applying the lensing estimator to these ILC maps we also inpaint SZ clusters as described for the ACT only maps.}

The maximum CMB multipole, $\ell_{\rm CMB}^{\rm max}=3000$, typically used in CMB lensing analysis is
motivated by the desire to reduce contamination from foregrounds such as the tSZ.
Since the tSZ bias is nulled in this new estimator, it is plausible
that this maximum multipole is unnecessarily conservative and can be increased, thus
improving the signal-to-noise of the estimator. Motivated by this possibility, we
increase our maximum multipole for the tSZ-free TT estimator map somewhat, to
$\ell_{\rm CMB}^{\rm max}=3350$; we perform a null test (see next section) to test for problematic contamination from other foregrounds such as CIB or kSZ. (This type of contamination becomes large when we use a higher lmax, such as 3500 and 4000, causing null test failures; for this reason, we choose to only modestly increase $\ell_{\rm CMB}^{\rm max}$ to 3350.) Furthermore, since the ILC maps include information from
\emph{Planck} for $\ell<500$, we also relax the minimum multipole cut from $\ell_{\rm CMB}^{\rm min}=500$ to $\ell_{\rm CMB}^{\rm min}=100$, providing additional gains in signal-to-noise.

We then create a foreground-cleaned minimum variance lensing map as in Equation \ref{eq:mvlensingnomc}. The coadding procedure is the same as for the ACT-only lensing map, except that temperature lensing is now obtained from the tSZ-free symmetric estimator $\hat\kappa^{TT}_{\rm symm,
 ~tSZ-free}$. We successfully repeat the lensing validation described in Section 3 with our new foreground cleaned estimator; the results are shown in Figure \ref{fig:lensverification}.

\section{\label{sec:CrossCorr}Galaxy cross-correlation measurement}

In the previous sections, we have introduced two types of CMB lensing maps, which
will be publicly available as part of the upcoming data release DR4 associated with \cite{Aiola:2019} and \cite{Choi:2019}. As an example of their utility, we cross-correlate these lensing maps with galaxies from the BOSS survey's CMASS galaxy catalog.

\subsection{The CMASS galaxy map}

We use the CMASS galaxy catalog (with redshifts $z \in [0.43, 0.7]$) provided by the DR12 release of the BOSS spectroscopic survey  \footnote{\url{http://www.sdss3.org/surveys/boss.php}} to construct a galaxy overdensity map. 
Given a pixel $\vec{x}$, we estimate the galaxy overdensity as 
\begin{equation}
\delta_g(\vec{x})= \frac{\sum_{i\in {\rm unmasked}~ \vec{x}} w_i}{\frac{1}{N}\sum_{i,{\rm all~unmasked}} w_i}-1
\end{equation}
where $N$ is the number of unmasked pixels (see below) and following \cite{Pullen:2016,Miyatake:2017} each galaxy $i$ inside the pixel $\vec{x}$ is weighted according to
\begin{equation}
w = (w_{\rm noz}+w_{\rm cp}-1)w_{\rm see}w_{\rm star}
\end{equation}
where $w_{\rm noz}$ accounts for redshift failures, $w_{\rm cp}$ for fiber
collisions, $w_{\rm star}$ for bright star contamination and $w_{\rm see}$ for effects of seeing. 

The galaxy mask used to mask pixels is created using `random catalogs' provided by the BOSS
collaboration; these catalogs contain a dense sampling of sky locations proportional to the survey conditions but not to any cosmological galaxy clustering signal.  The random catalogs are mapped to a number density count
map (created setting $w=1$) and then smoothed with a Gaussian beam with a width
corresponding to a standard deviation of 2 arcminutes. To obtain the final mask, we then set to zero the regions of the smoothed randoms' counts below a threshold of $10^{-3}$. The above choices are made so as to preserve survey information without picking up fluctuations in the random sampling. Our baseline analysis accounts for the effect of this mask simply by applying an overall scaling factor which compensates for the loss in power due to zeroed regions, as described in the next section. In general, the mask can also cause coupling of Fourier modes of the map leading to a modification of the estimated power spectrum. Although these effects are expected to be small since our mask is smooth, we test the impact of the mask on our cross-correlation measurement.

\begin{figure}
    \centering
    \includegraphics[width=9cm]{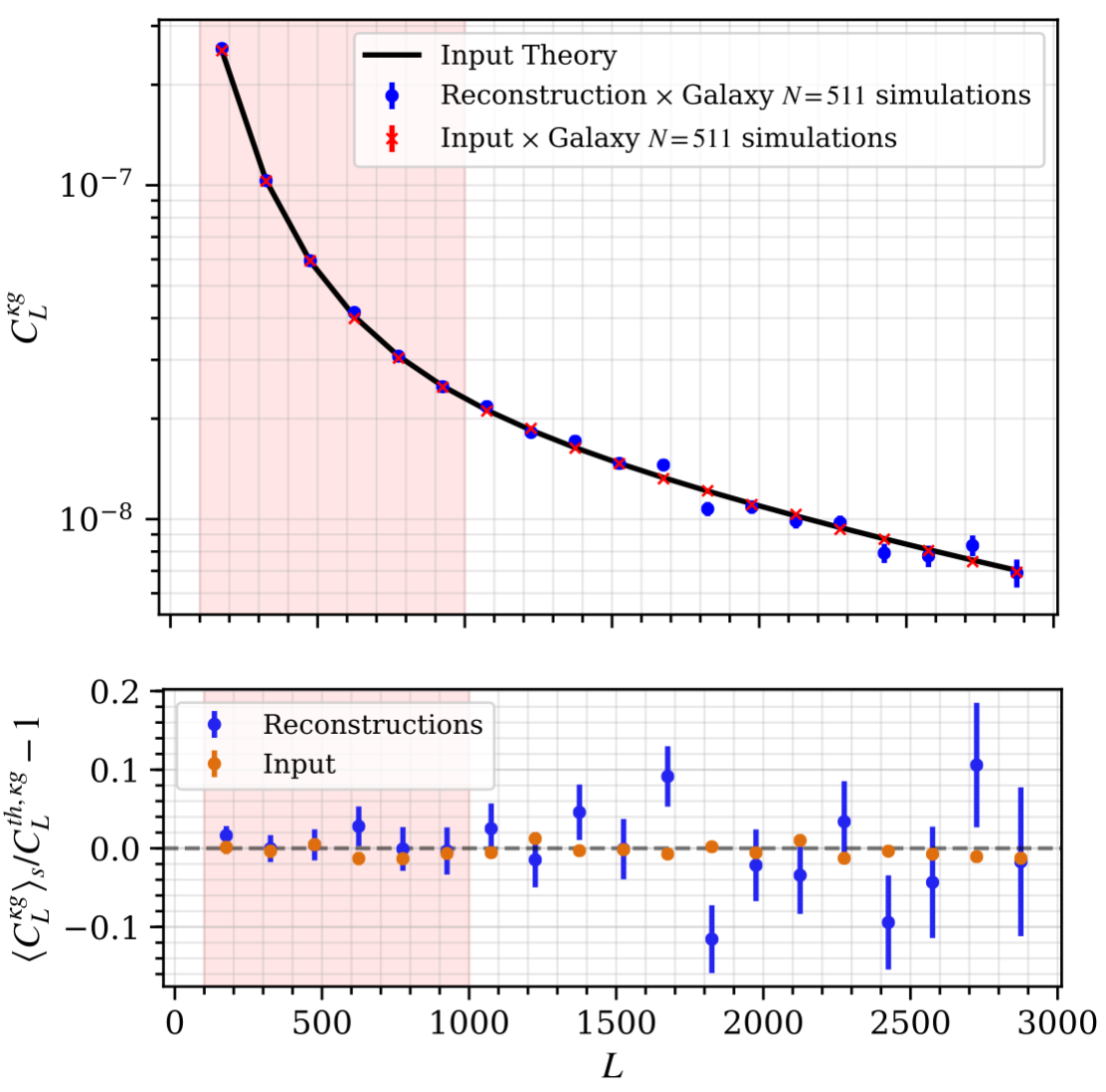}
    \caption{Lensing reconstruction test, as for Figure 1, but now correlating with a simulated galaxy field instead of the input lensing convergence field. The shaded region shows the multipole range used for the cosmological analysis. The lower panel shows the fractional difference with respect to the theory curve. (The {\tt BN} patch gives similar results.)}
    \label{fig:maskverification}
\end{figure}

We validate the treatment of the galaxy mask by applying it to mock Gaussian galaxy overdensity simulations which are correlated with the lensing signal according to a theoretical cross-spectrum with a fiducial bias $b=2$. We verify that the cross-power spectrum measured from these simulations, with a multiplicative correction for the mask as described in the next section, reproduces
the original input theory cross-correlation signal. As shown in Figure
\ref{fig:maskverification} we recover $C_l^{\kappa_s g}$ to better than five percent over the cosmological analysis range, with no indication of an overall bias.

{}

\subsection{Extracting power spectra and obtaining the covariance matrix}

Having constructed CMB lensing and galaxy maps we measure their cross-power spectra. Binned cross-power spectrum measurements are obtained using the following estimator valid for statistically isotropic fields,

\begin{equation}
\hat{C}^{\kappa g}_{L_b}=\frac{1}{w^{\kappa g}} \frac{1}{N_{\mathcal{A}}}\sum_{\Lb\in \mathcal{A}} \kappa_{\Lb}^{obs}g^{obs*}_{\Lb}
\label{eq:caverage}
\end{equation}
where $\mathcal{A}$ is an annulus in the Fourier plane with average radius $L =|\Lb|$, $N_{\mathcal{A}}$ gives the number of modes in this annulus, and $w^{\kappa g}$ is a correction factor due to masking that depends on the masked
fields taken in consideration. For a slowly varying window function this is given by

\begin{equation}
    w^{\kappa g}=\langle W_\kappa^{2}(\vec{x})W_g(\vec{x})\rangle
\end{equation}

\noindent where $W_\kappa$ is the mask we apply to our CMB map before lensing reconstruction and $W_g$ is the mask applied to our galaxy overdensity map. Two powers of the CMB mask appear in the correction above because the lensing reconstruction is a quadratic estimator involving two powers of the CMB map.\footnote{To avoid confirmation bias we did not plot a y-axis scale or overplot a theory curve over our cross-spectrum measurement until all the null tests and systematics checks, described in Section 6, had been successfully passed.}

We obtain the covariance matrix for the cross-spectra from simulations as follows:
\begin{equation}
     \hat{C}_{L_b,L_b'} = \langle ({C}_{L_b}^S-\langle \hat{C}_{L_b}^S \rangle_S)({C}_{L_b'}^S-\langle \hat{C}_{L_b'}^S \rangle_S)^T \rangle_S
\end{equation}
where the column power spectrum vector is $\hat{C}^S = (C_{L_b}^{\kappa_S g_S})^T$
and the average is over the simulations $S$.

To calculate this matrix, we cross correlate the $N=511$ lensing reconstruction simulations with the QPM mock catalogs of CMASS galaxies \citep{QPMMocks}. The cosmological signals in these simulations and catalogs are uncorrelated. We expect this not to be problematic because the uncorrelated part of the cross-correlation error dominates over the sample variance contribution. We verify this by calculating Gaussian theory standard errors with and without the $\left(C_{L_b}^{\kappa g}\right)^2$ sample variance term that arises from the presence of correlated structures, finding sub-percent level agreement between the two calculations.

The inverse covariance matrix obtained from $N$ simulations is calculated as in \cite{Hartlap}:
\begin{equation}
    \widehat{C^{-1}} = \beta \hat{C}^{-1} \label{eq:invmatrix}
\end{equation}
where $\beta = \frac{N-p-2}{N-1}$ with $p$ the number of angular bins.

Finally, we note that some care is required when choosing the range of scales
$L_{\rm min} < L < L_{\rm max}$ which we use in our analysis. Our
theoretical model is expected to break down on smaller scales, since we are
assuming a simple scale-independent linear galaxy bias, ignoring baryonic feedback on the matter power spectrum and also assuming that
the non-linear matter power spectrum derived from HMCode \citep[][]{Mead}, implemented in CAMB, is reliable. We therefore initially pick a range of scales based on the cross-correlation measurement; we set the requirement that the difference between a linear theory calculation of the cross-spectrum and the non-linear (HMCode) calculation should not be larger than the $1$-$\sigma$ uncertainty for our cross-spectrum measurement. In this way, we obtain that the appropriate cutoff is approximately $L_{{\rm max}, \kappa g} = 1000$.

In addition to the small-scale cuts described above, we also wish to avoid systematic errors which enter on large, degree-angular scales. On the galaxy side, such systematic errors include depth and selection function variations over the survey footprint; on the CMB lensing side, the main large-scale limitation is the challenge in simulating and subtracting the mean-field term sufficiently accurately, since it grows rapidly towards very low $L$ ($L<50$). While many systematics are nulled in cross-correlation, they could induce additional variance, and to be safe we choose $L_{\rm min}=100$ for our analysis; at this scale, the power spectrum of the mean field is still smaller than that of the signal.

For our measurements, we choose a binning of $
\Delta L = 150$; with this binning, we find that the correlations between different bandpowers are not strong ($<13\%$).

\subsection{Galaxy cross-correlation: results}
\begin{figure}
    \centering
    \includegraphics[width=9cm]{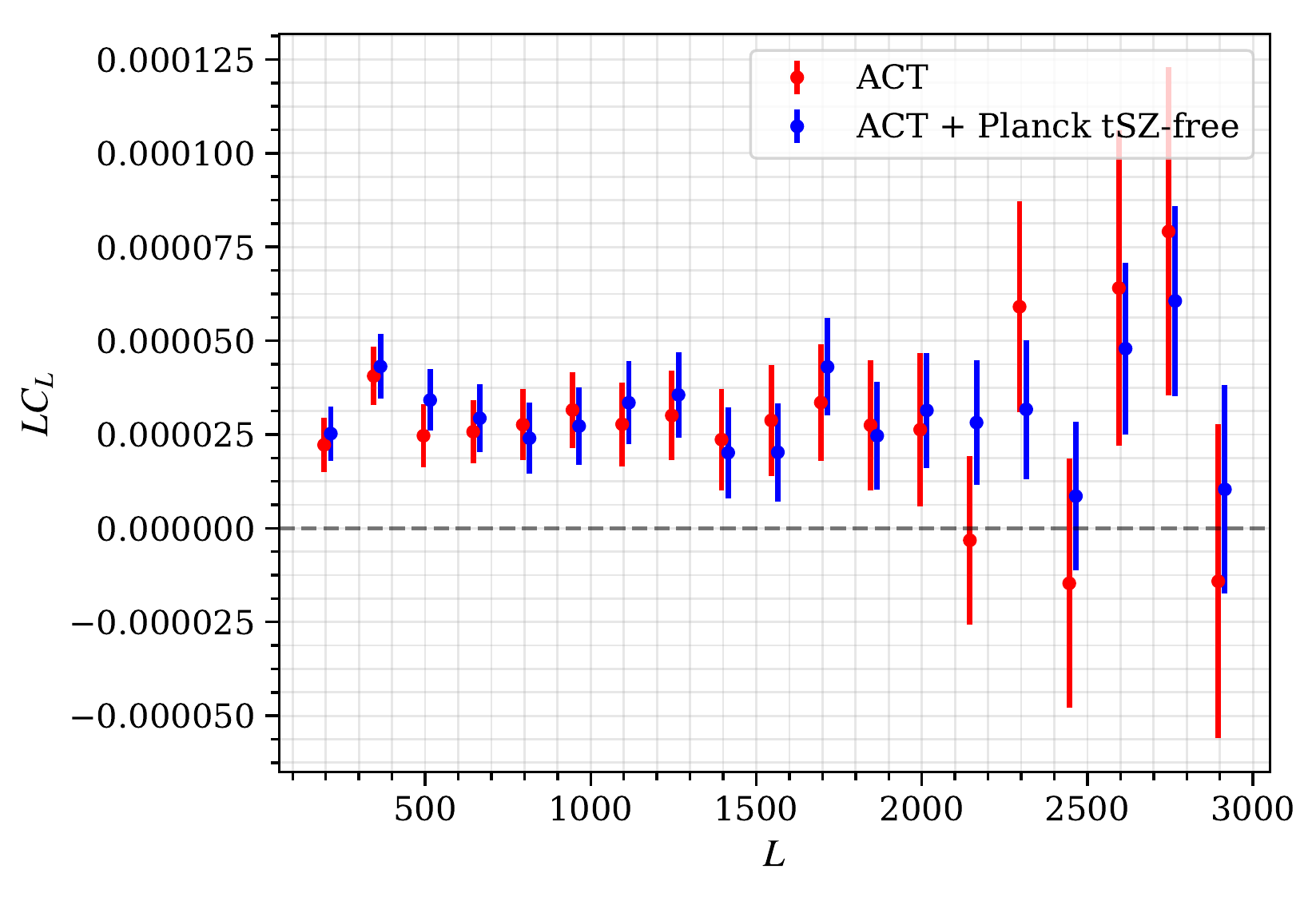}
    \caption{The cross-correlation between CMASS galaxies and CMB lensing convergence reconstructions from ACT. The cross-correlation measurements in the {\tt D56} and {\tt BN} patches are coadded to obtain these results. The red points show results using an ACT-only lensing map, the blue points show results using a lensing map that has been tSZ cleaned. The multipole values for different versions of lensing maps are slightly offset for visualisation purposes. See Figure \ref{fig:cosmoanalysis} for comparison with a theory curve fit to the cleaned measurement.}
    \label{fig:crosscorrmes}
\end{figure}

In Figure \ref{fig:crosscorrmes} we show the new tSZ-free CMB lensing -- galaxy cross-correlation measurement. We also show
the same cross-correlation with the ACT-only lensing maps, which have not been cleaned of tSZ. 

{}

A small shift between the bandpowers can be seen. It appears to match the form expected from bias due to tSZ in the ACT-only maps, {\it i.e.,} a deficit on large scales and an excess on
small scales. However, the difference was not found to deviate from zero by a statistically significant amount, with a $\chi^2$ probability to exceed (PTE) of 0.29 (for the cosmology range). Nevertheless, we note that the difference is a good fit to a simplified foreground bias model (given by a 10\% deficit in the cross-correlation at $L<800$); the $\chi^2$ to this model is lower than for a fit to null by $\Delta \chi^2 = 2.2$.

\begin{figure*}
    \centering
    \includegraphics[width=12cm]{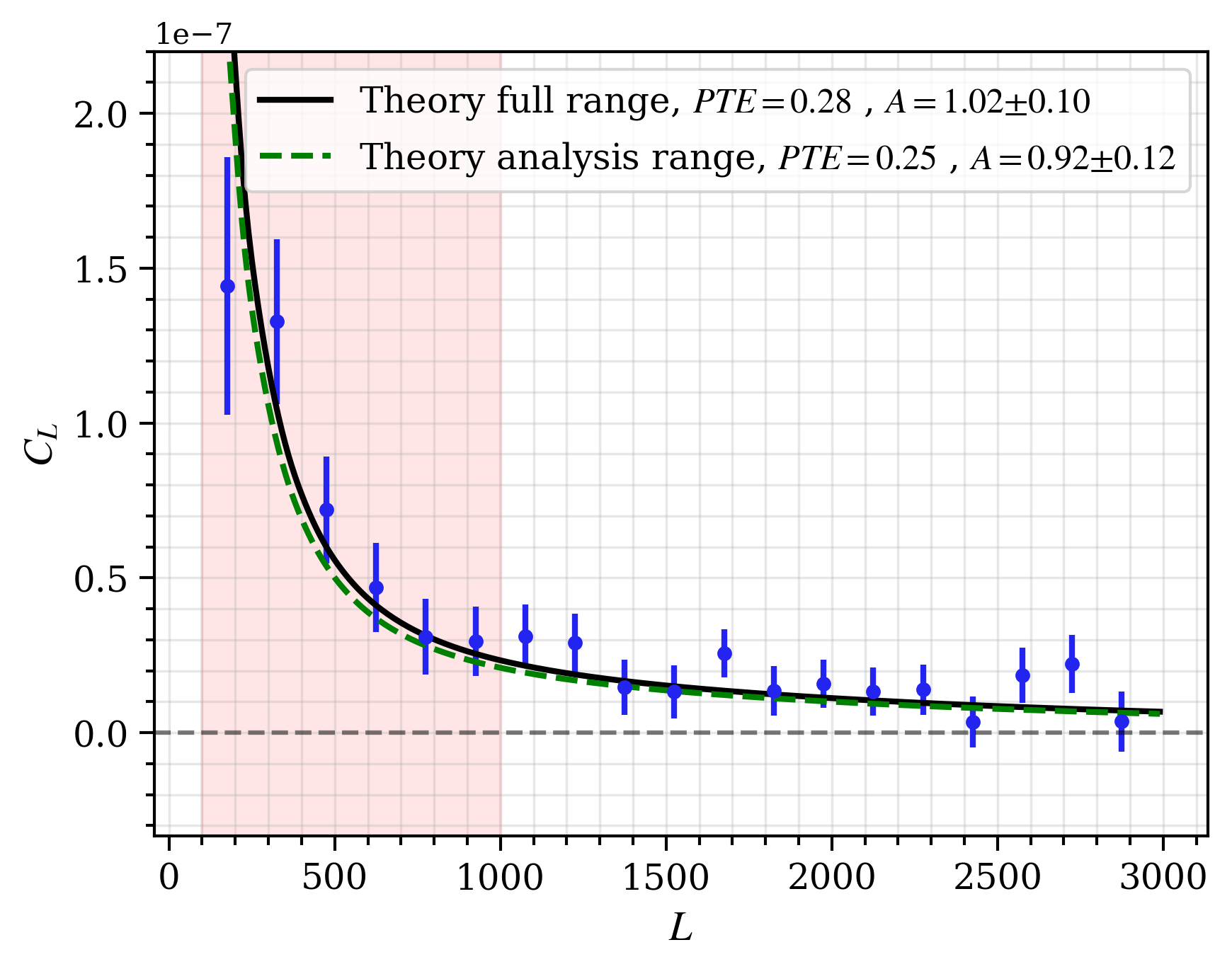} 
    \caption{This plot shows our main CMB lensing -- BOSS galaxy cross-correlation measurement with the ACT+\emph{Planck} tSZ free lensing maps (blue points). A \emph{Planck}-cosmology (and fiducial galaxy bias $b_{\rm{fid}}=2$) theory template, with a free amplitude fit to the data ($A=b/b_{\rm{fid}}$), is also indicated with a dashed line. The green dashed theory curve is fit only over a restricted analysis range (shaded region for scales $100<L <1000$); the black solid curve is fit over the full $L$ range shown in this plot. (The bandpowers are nearly independent, with the off-diagonal elements of the covariance matrix showing correlations of less than $13 \%$). We find good consistency in both cases with the \emph{Planck}-cosmology derived theory template.}
    \label{fig:cosmoanalysis}
\end{figure*}{}

Although the tSZ-free measurement contains no bias from tSZ, the measurement errors on large scales are similar, which highlights the power of this new technique in
providing unbiased measurements that do not sacrifice significant signal-to-noise.\footnote{The fact that measurement uncertainties do not significantly increase in our method, although it removes foregrounds, is not just due to the inclusion of \emph{Planck} data; indeed, a naive application of the standard quadratic estimator to tSZ-deprojected ACT+\emph{Planck} maps gives cross-correlation uncertainties that are  $\approx$50\% larger. \emph{Planck} enables better multifrequency cleaning, rather than adding much raw statistical weight to the ACT maps.}

We adopt the tSZ-cleaned cross-correlation as our standard analysis. We fit the cross-correlation with a fiducial theory model; this model uses both fiducial \emph{Planck} parameters as well as a fiducial linear bias of $b=2$, motivated by previous BOSS analyses \citep{Alam}. The cross-correlation measurement as well as a fit of the amplitude of this fiducial model are shown in Figure \ref{fig:cosmoanalysis}. It can be seen that, for both the restricted analysis multipole range and the full range, the amplitudes obtained are consistent with the fiducial value ($A=1$). In particular, we obtain $A=0.92\pm 0.12$ for a fit to the restricted analysis range and $A=1.02\pm 0.10$ for the fit to the full range of scales. Both theory curves are a good fit to the measurements, with $\chi^2$ PTEs of $0.25$ and $0.28$ respectively. Thus, we find good consistency in both cases with the \emph{Planck}-cosmology derived theory template.

%______________________________________________________________

%______________________________________________________________

\section{Systematics and Validation of the cross-correlation measurement\label{sec:systematics}}

We perform several tests for systematic errors to validate both our lensing maps and our cross-correlation measurement. Note that the relevant covariance matrices are obtained from Monte Carlo simulations of each test. These covariances are used to derive a chi-squared to null probability to exceed (PTE) for every test.

\begin{figure}
    \centering
    \includegraphics[width=9cm]{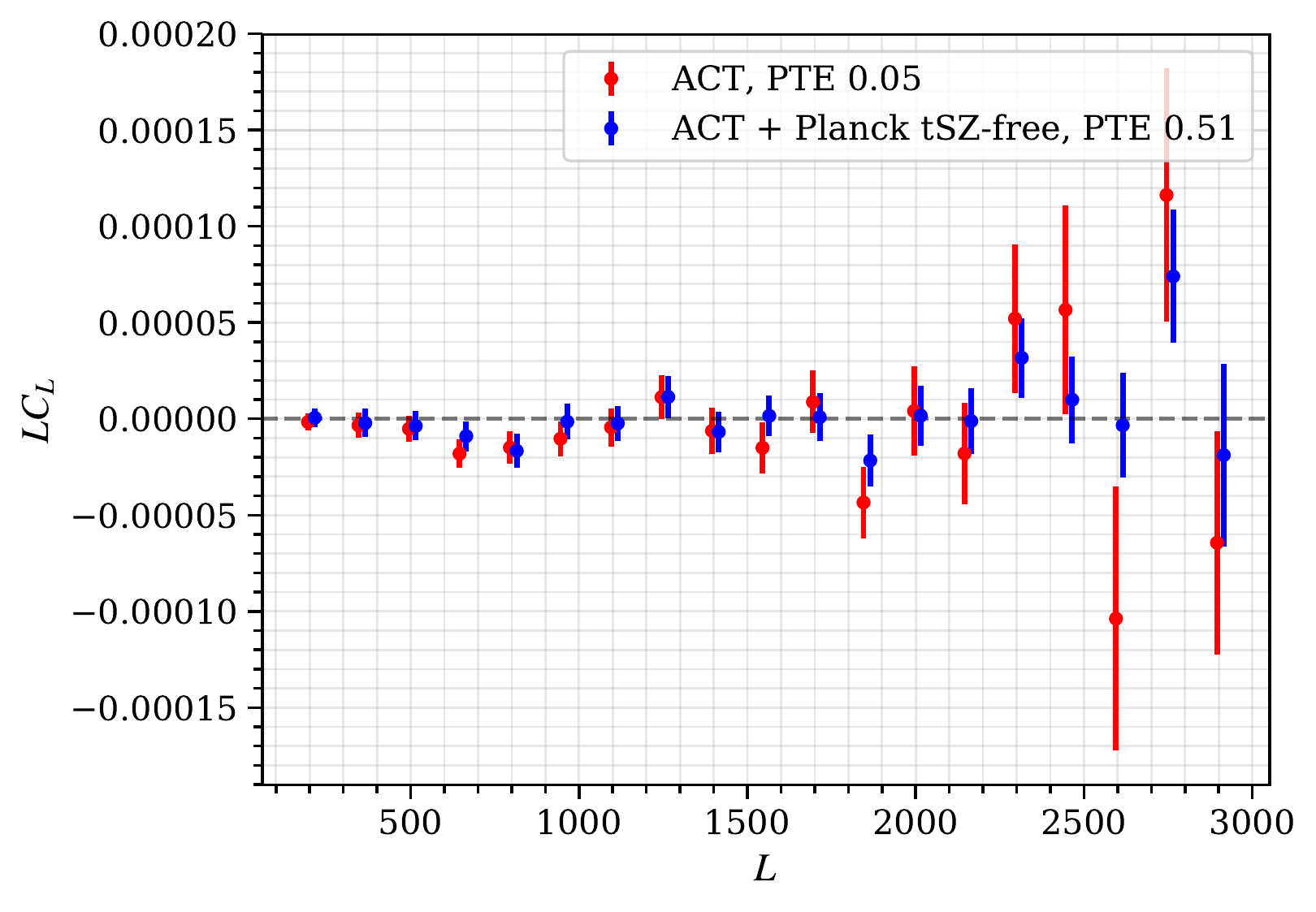} 
    \caption{A curl null test: verification that the extracted curl-lensing -- galaxy cross-correlation, which should be negligibly small in the absence of systematic errors, is consistent with the null hypothesis. The results shown are for a combination of both {\tt D56} and {\tt BN} patches. The $\chi^2$ probability-to-exceed (PTE) for this null test is also shown in the legend.}
    \label{fig:omegaverification}
\end{figure}

Our first null test relies on the fact that we expect the cosmological lensing signal
from gravitational scalar perturbations to give rise to gradient-like
deflections. Hence, this deflection field should be irrotational, with zero
curl.\footnote{The potential cosmological curl signal coming from tensor
  perturbations at linear order or from scalar perturbations at second order is
  well below current sensitivity.} In contrast, systematics that mimic lensing
can have non-zero curl. Therefore, a detection of a curl signal can be a
signature of unknown systematic errors present in our data. By using a quadratic estimator $\hat{\Omega}^{XY}(\Lb)$ similar to that for the lensing potential but with different filters
\citep{cooray:2005} (essentially the dot product in the potential estimator
is replaced by a cross product), it is possible to extract the curl signal and cross correlate it with the BOSS galaxy field. As shown in Figure \ref{fig:omegaverification}, this cross-correlation signal is consistent with zero, with a PTE of $0.51$ for the tSZ-cleaned lensing cross-correlation. We note that for the ACT-only cross-correlation, the PTE is only $0.05$, although this may simply be due to a statistical fluctuation.

{}

\begin{figure}
    \centering
    \includegraphics[width=9cm]{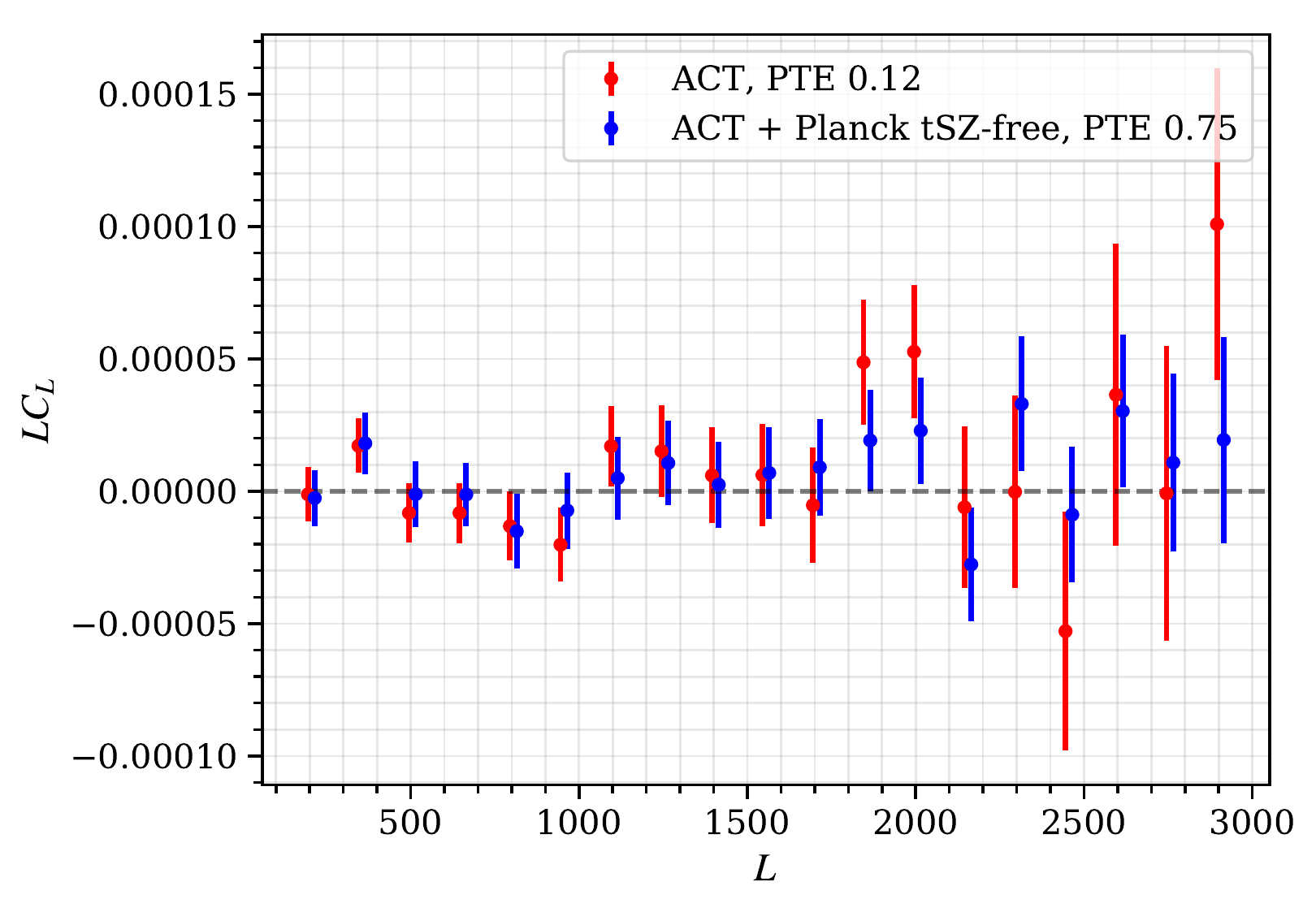}
    \caption{A null test verifying that cross-correlating the lensing map on one field with the galaxy map on the other field (and combining both spectra) gives a signal consistent with zero. Red points show ACT-only results, blue points show tSZ-deprojected lensing results. The fact that the PTEs in both cases are consistent with zero signal supports the conclusion that our uncertainty calculations are correct.}
    \label{fig:independentverification}
\end{figure}{}

\begin{figure}
    \centering
    \includegraphics[width=9cm]{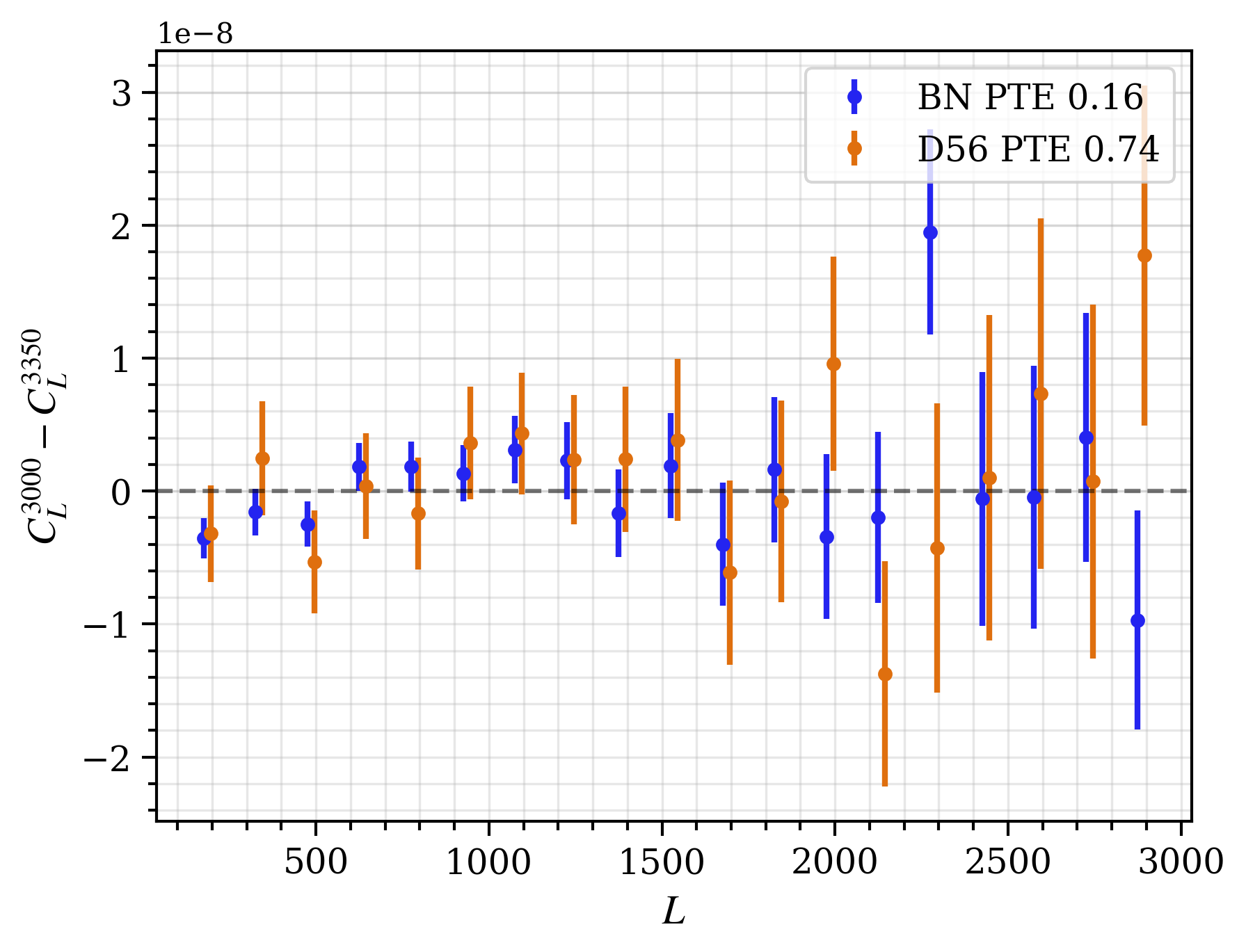} 
    \caption{An extragalactic foreground null-test for the cleaned maps. We show the difference between the cross-correlations of CMASS with the tSZ-deprojected lensing maps for the cases of $l_{\rm CMB,max}=3000$ and $l_{\rm CMB,max}=3350$, where $l_{\rm CMB,max}$ is the maximum CMB multipole used in the lensing reconstruction. Since extragalactic foregrounds rise rapidly towards high $l$, a substantial foreground residual in the cross-correlation would cause a null test failure. However, our null test results shown here are consistent with zero contamination for both fields (blue: points for {\tt BN}, orange: points for {\tt D56}).}
    \label{fig:tSZverification_notsz}
\end{figure}{}

\begin{figure}
    \centering
    \includegraphics[width=9cm]{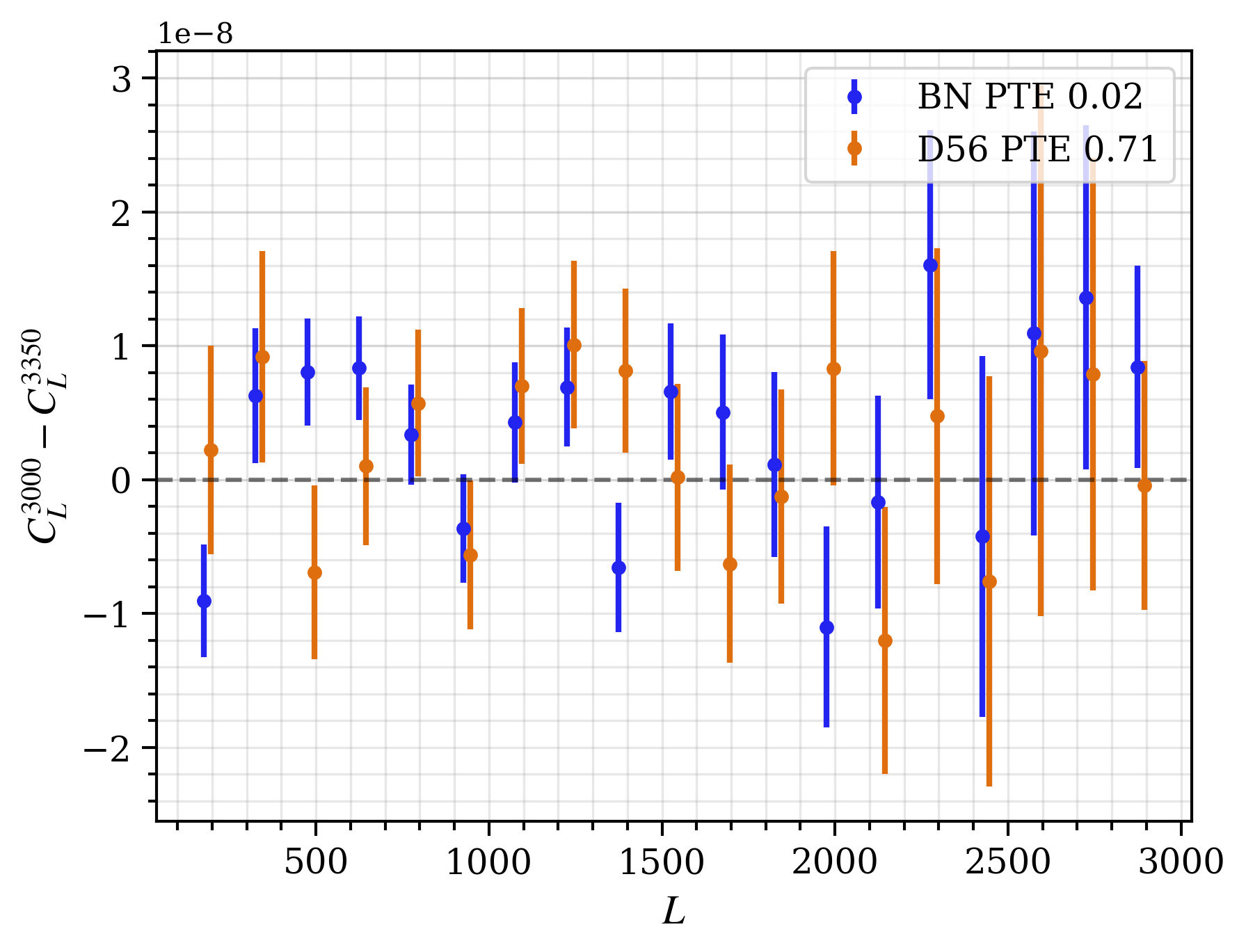} 
    \caption{The same test as shown in Figure \ref{fig:tSZverification_notsz}, but applied to the ACT-only maps which have not been foreground cleaned (blue: points for {\tt BN}, orange: points for {\tt D56}). The PTE for {\tt BN} shows a (mild) failure of the null test, as is expected if foreground residuals are important.}
    \label{fig:tSZverification_kspace}
\end{figure}{}

As a second test, we cross-correlate the galaxy map of one patch with the
lensing convergence map of the other patch\footnote{To perform this correlation, we
  extend with zero values the maps of the smaller patch, in this case {\tt D56},
  so that the two fields have the same size.} and check for consistency with
zero. It is very difficult to imagine systematics that would correlate fields
that are so far apart, and so this test primarily serves as a validation of our covariance
matrix and uncertainty calculation. In Figure  \ref{fig:independentverification}
we see that the results of this null test are consistent with zero, with a PTE of $0.75$ obtained for the tSZ-cleaned lensing map and $0.12$ for the ACT-only map.

Thirdly, we wish to test for the presence of residual foreground-induced bias
in the cross-correlation measurement, even though we
expect to be insensitive to the dominant tSZ contamination when using our
symmetric cleaned lensing estimator. To test for residual foreground biases from the CIB, kSZ \citep[e.g.][]{kSZBias} or other sources (including those arising from incomplete tSZ cleaning),
we make use of the fact that foreground contamination should become worse as the maximum CMB multipole 
$\ell_{\rm CMB, max}$ used in the lensing reconstruction increases. If our
foreground cleaning is working as expected and residual foregrounds are negligible, results with a high $\ell_{\rm CMB,
  max, high}$ and a lower $\ell_{\rm CMB, max, low}$ used in the reconstruction
should be consistent. In Figure \ref{fig:tSZverification_notsz} we show this
foreground null test for the symmetric cleaned estimator; in particular, we plot the difference $C_{L}^{\kappa_{\rm low}g}-C_{L}^{\kappa_{\rm high}g}$
of the cross-correlation $C_{L}^{\kappa_{\rm high}g}$ with a higher lensing reconstruction $\ell_{\rm CMB, max, high}=3350$ (the baseline used in this work) and the cross-correlation with a lower $\ell_{\rm CMB, max, high} = 3000$,
$C_{L}^{\kappa_{\rm low}g}$. It can be seen that this difference is consistent with
zero overall, with PTEs of $0.74$ and $0.16$ found for {\tt D56} and {\tt BN} respectively. The error bars are obtained from simulations and hence take
into account the covariance between the two spectra. For comparison, in Figure
\ref{fig:tSZverification_kspace} we perform the same test for the ACT-only maps which are not free of tSZ; perhaps unsurprisingly, we find a (mild) null test failure (PTE of $0.02$) for the {\tt BN} patch, although the {\tt D56} PTE of $0.71$ still appears acceptable.

Finally, to check for sensitivity to large-scale systematics, we vary the lowest multipole $L_{\rm min}$ of the first bandpower of the cross-correlation measurement; we find that the value of the first bandpower is stable. This was the only null test done after we unblinded.

Our suite of null tests does not show evidence for foreground or systematic
contamination to our measurement, as long as we use the symmetric cleaned lensing estimator. In particular, for the combined {\tt BN}+{\tt D56} cleaned measurement we find a PTE of 0.28 for the foreground residual test, showing no evidence for foreground contamination in the cross correlation.

%______________________________________________________________

\section{\label{sec:Discussion}Discussion}

In this paper, we present maps of CMB lensing convergence derived from ACT
observations made in 2014-15. The lensing maps are constructed in two different
ways: first, by applying the standard quadratic lensing estimator to only ACTPol
CMB data; second, by implementing a new ``symmetric'' foreground-cleaned lensing estimator, which makes use of component separated ACTPol+\emph{Planck} CMB maps to return lensing maps that are free of tSZ-bias in cross-correlation.

We report combined cross-correlation measurements of our CMB lensing maps with BOSS CMASS galaxies at $\approx 10\sigma$ significance. We find that the use of our new tSZ-free estimator does not significantly increase the size of measurement uncertainties.

We will release these lensing maps to enable other cross-correlation analyses
with large-scale-structure. However, several caveats should be kept in mind when
making use of these maps. Only the bispectrum $\langle g
T_{\rm{tSZ}}T_{\rm{tSZ}} \rangle$ tSZ contamination is nulled in our procedure,
where $T_{\rm{tSZ}}$ is the tSZ signal and $g$ is the large-scale structure
field (e.g., galaxy overdensity or galaxy shear); this is the dominant source of
contamination for near-term cross-correlations with $z<1$ structure. Users of
these maps should be aware that high-redshift cross-correlations can be
contaminated with the CIB field $T_{\rm{CIB}}$, both through $\langle
gT_{\rm{CIB}} T_{\rm{CIB}}  \rangle$ as well as through its correlation with the
tSZ $\langle gT_{\rm{tSZ}} T_{\rm{CIB}}  \rangle$. For cross-correlations where
CIB contamination is more of a concern than tSZ contamination (e.g., for
cross-correlations with the CIB itself), our pipeline allows the application of
the analog of our symmetric cleaned estimator on CIB-deprojected maps from
\cite{Madhavacheril:2019}. Such analyses should be validated on realistic
simulations \citep[e.g.,][]{0908.0540,2001.08787} to verify that the tSZ
contamination is sub-dominant. Looking beyond the 2014 and 2015 data used in
this work, high-resolution 230 GHz data collected with the Advanced ACTPol
instrument from 2016 and onward should allow for simultaneous deprojection of
both the tSZ and CIB contamination for use in symmetric cleaned estimators that are robust at all redshifts. The contamination from the kSZ will, however, remain, since the kSZ has the same blackbody frequency spectrum as the primary CMB, although the contamination is much lower in amplitude \citep{DasForegrounds, kSZBias}. Alternatives to our method include shear-only reconstruction \citep{1804.06403} (which requires the inclusion of smaller scales in the CMB map to achieve similar signal-to-noise) and source hardening \citep{Osborne} (primarily targeted at reducing contamination from point sources and clusters). The optimal combination of all of these methods that minimizes bias (both from foregrounds and higher-order effects) and maximizes signal-to-noise remains an open problem.

We also caution users that the auto-spectrum of the lensing potential presents a much broader set of analysis challenges, both for mitigation of foregrounds (where the CIB contamination is expected to be larger \citep{vanEngelenForegrounds}) and for characterization and subtraction of reconstruction noise bias. The latter requires an extensive set of simulations  \citep[e.g.,][]{Sherwin:2017,1412.4760} and methods robust to mismatch of simulations and the observed sky \citep[e.g.,][]{Namikawa:2013}. The CMB lensing auto-spectrum from ACT data from 2014 and 2015 will appear in a separate work. In addition, care should be taken when attempting to interpret the signal from stacking massive clusters on our released CMB lensing maps; first, because inpainting and masking steps can introduce complications, and second, because higher order effects can bias the standard quadratic estimator near the most massive clusters \citep{0701276}.

This work lays the foundation for upcoming, higher precision ACTPol and Advanced ACT cross-correlations with galaxy and lensing surveys. For upcoming cross-correlation analyses with ACT and other experiments, powerful methods to obtain foreground free measurements are necessary; our work represents one promising solution to this problem. 

\section*{Acknowledgements}
We thank Shirley Ho, Alex Krolewski, Will Percival and Anthony Pullen for useful discussions regarding the BOSS survey. Some of the results in this paper have been derived using the healpy~\citep{Healpix1} and HEALPix~\citep{Healpix2} package. This research made use of Astropy,\footnote{http://www.astropy.org} a community-developed core Python package for Astronomy \citep{astropy:2013, astropy:2018}. We also acknowledge use of the matplotlib~\citep{Hunter:2007} package and the Python Image Library for producing plots in this paper, and use of the Boltzmann code CAMB~\citep{CAMB} for calculating theory spectra. 

This work was supported by the U.S. National Science Foundation through awards AST-1440226, AST0965625 and AST-0408698 for the ACT project, as well as awards PHY-1214379 and PHY-0855887. Funding was also provided by Princeton University, the University of Pennsylvania, and a Canada Foundation for Innovation (CFI) award to UBC. ACT operates in the Parque Astron\'{o}mico Atacama in northern Chile under the auspices of the Comisi\'{o}n Nacional de Investigaci\'{o}n Cient\'{i}fica y Tecnol\'{o}gica de Chile (CONICYT). Computations were performed on the GPC and \emph{Niagara} supercomputers at the SciNet HPC Consortium. SciNet is funded
by the CFI under the auspices of Compute Canada, the Government of Ontario, the Ontario Research Fund
-- Research Excellence; and the University of Toronto. The development of multichroic detectors and lenses was supported by NASA grants NNX13AE56G and NNX14AB58G.   Colleagues at AstroNorte and RadioSky provide logistical support and keep operations in Chile running smoothly. We also thank the Mishrahi Fund and the Wilkinson Fund for their generous support of the project.

OD, BDS, FQ and TN acknowledge support from an Isaac Newton Trust Early Career Grant and from the European Research Council (ERC) under the European Unions Horizon 2020 research and innovation programme (Grant agreement No. 851274). BDS further acknowledges support from an STFC Ernest Rutherford Fellowship. MSM acknowledges support from NSF grant AST-1814971. NB acknowledges support from NSF grant AST-1910021. EC is supported by a STFC Ernest Rutherford Fellowship ST/M004856/2 and the STFC Consolidated Grant ST/S00033X/1. R.D. thanks CONICYT for grant BASAL CATA AFB-170002. DH, AM, and NS acknowledge support from NSF grant numbers AST-1513618 and AST-1907657. JCH acknowledges support from the Simons Foundation and the W.~M.~Keck Foundation Fund at the Institute for Advanced Study. Flatiron Institute is supported by the Simons Foundation. MHi. acknowledges financial support from the National Research Foundation (NRF) South Africa. LM received funding from CONICYT FONDECYT grant 3170846. KM acknowledges support from the National Research Foundation of South Africa.

%______________________________________________________________

\appendix

\section{\label{sec:mapProcessing} CMB Map Pre-processing for Lensing reconstruction}

In this appendix, we describe in more detail the preprocessing of the ACT CMB maps which are used in the lensing reconstruction process.

The ACT raw maps are made available as four map splits $D_{A, f,j}, j\in \{1,2,3,4\}$ with the same signal but independent instrumental noise contributions through the time-interleaved splitting scheme described in \cite{Aiola:2019} and \cite{Choi:2019}, for each frequency $f$ and instrumental array $A$. For the {\tt D56} region, data are from seasons 2014
and 2015 and observations of the sky are made from the following combinations of
array-frequency $(A,f)$: (PA1-2014, 150), (PA2-2014, 150), (PA1-2015, 150), (PA2-2015, 150), (PA3-2015, 150), (PA3-2015,98), where only the dichroic PA3 array includes observations at both 98
GHz and 150 GHz. For  the {\tt BN} region, the data are from season 2015 only, for the
combinations $(A,f)$: (PA1-2015, 150), (PA2-2015, 150), (PA3-2015, 150) and (PA3-2015,98). Here, (PA3-2015,150), for example, corresponds to a map made using measurements from the 150 GHz channel of the PA3 detector array collected during the 2015 observing season.

The temperature maps that enter the ACT+\emph{Planck} tSZ-free lensing maps are pre-processed and coadded (with appropriate tSZ deprojection) as described in \cite{Madhavacheril:2019}. All other maps (i.e. temperature maps for the ACT-only lensing maps and the polarization maps) are pre-processed and co-added as follows:
\begin{enumerate}
    \item  To reduce noise and bias from radio sources and to make subsequent Fourier transforms well-behaved, we use source subtracted maps (see \cite{Choi:2019,Aiola:2019}). Some residuals are left in these at the locations of bright compact sources; these are in-painted within each split using the catalog and  maximum-likelihood method described in \cite{Madhavacheril:2019}, {\it i.e.,} we fill holes around compact sources with a constrained Gaussian realization. These holes of 6 arcminute radius are inpainted jointly for T, Q, U. The algorithm used follows the brute-force approach presented in \cite{Bucher:2012}.
    We then use these splits to obtain a co-added map $D_{A, f}$ using maps of the inverse
    white-noise variance in each pixel as well as two sub-splits $D_{A, f, 1}=\sum_{j=1,2}D_{A, f,j}$
    and $D_{A, f, 2}=\sum_{j=3,4}D_{A, f,j}$ with independent noise. We use
    these two sub-splits to obtain an estimate of the 2d Fourier space noise
    power spectrum $N_{A, f}(\ellb)$, by taking the difference between the mean
    auto-spectrum of each sub-split and the mean cross-spectrum between the
    sub-splits, and subsequently smoothing it.\\
    \item We apply an apodized mask to each map which restricts our
      analysis to the well-crosslinked region used for power spectrum
      measurements in \cite{Choi:2019,Aiola:2019}. To account for pixelization
      effects, we deconvolve the pixel window function from each map in 2D
      Fourier space.  \\
    \item We next combine the various maps $D_{A, f}$ into a single CMB map $M$ on
      which the lensing reconstruction is performed, for each of T, Q and U.  Unlike in previous work where a real-space coaddition was used
      \citep{Sherwin:2017}, we now co-add
      the maps in 2D Fourier space (since this is more optimal for multifrequency data with different beams) as follows: $M(\ellb)=B_{A_{c}, f_{c}}(l)\sum_{(A,f)}w_{A,
        f}(\ellb)D_{A,f}(\ellb) B_{A,f}^{-1}(\ell)$ where
      \be
      w_{A, f}=\frac{N_{A, f}(\ellb)B_{A,f}^{-2}(\ell) }{ \sum_{(A,f)}
        N_{A, f}(\ellb)B_{A,f}^{-2}(\ell)}
      \ee
      are normalized inverse variance weights. We note that here a deconvolution of the
      harmonic space beam $B_{A,f}(\ell)$ is performed for each array, and finally a convolution to a common map beam $B_{A_{c}, f_{c}}(l)$ is reapplied; the choice of this beam does not matter since it is deconvolved later. This weighting scheme ignores correlations of the noise between arrays. Only
      the dichroic arrays (PA3,150 GHz) and (PA3, 98 GHz) have substantial
      ($\approx 40\%$) noise
      correlations on the scales considered in this work. While this choice of weighting is sub-optimal, on scales where the $(98-150) $ GHz
      correlation is important, our measurements are dominated by the CMB signal
      in the $98$ GHz frequency and thus neglecting these correlations will not substantially increase the
      lensing reconstruction noise. \\
    
    This procedure, performed separately for each of intensity $T$ and the $Q$
    and $U$ polarization stokes components, results in coadded CMB maps  $M_X$
    with  $X\in\{T, Q, U\}$. We repeat the same operations above on the
    sub-splits $D_{A, f, i} , i \in \{1, 2\}$ to obtain the corresponding maps
    $M_{X,i}, X\in\{T, Q, U\}$ from which we obtain an estimate of the
    experimental noise 2D power $N_X, X\in\{T, Q, U\}$ in the same way as
    described previously. These noise estimates of the co-added maps are used
    for optimal weighting in the lensing reconstruction. \\
    \item While the previously described inpainting procedure removes a large
      amount of radio source contamination, bright galaxy clusters show up in
      these maps as decrements due to the thermal Sunyaev-Zel'dovich
      effect. These add both noise and bias to the lensing estimation, and so we
      next in-paint a catalog of SZ clusters that have been internally
      detected. For this catalog, we use confirmed cluster locations
      inferred from co-add maps that include data up to the 2018
      season. 
      From this catalog, we select and inpaint all  the  clusters  with a signal-to-noise
      ratio greater than $5$.  The inpainting is performed (only in temperature) within circular holes
      of $5$-arcmin radii using the same method as for the compact sources. A small number of clusters near the edge of
      the mask that caused problems due to the discontinuous boundary were not inpainted. This is expected to have a negligible impact on our
      analysis as the number of such clusters is very small, with no
      particularly bright ones among them. After
      inpainting, we deconvolve by the common map beam chosen above.   
     \end{enumerate}
        The CMB temperature and polarization maps that result from these steps are used (following filtering and $E-B$ decomposition) as inputs to our lensing reconstruction pipeline, described in detail in Section 3.

\section{\label{sec:SZ} Noise properties of the symmetric foreground-cleaned estimator}

\begin{figure}
    \centering
    \includegraphics[width=9cm]{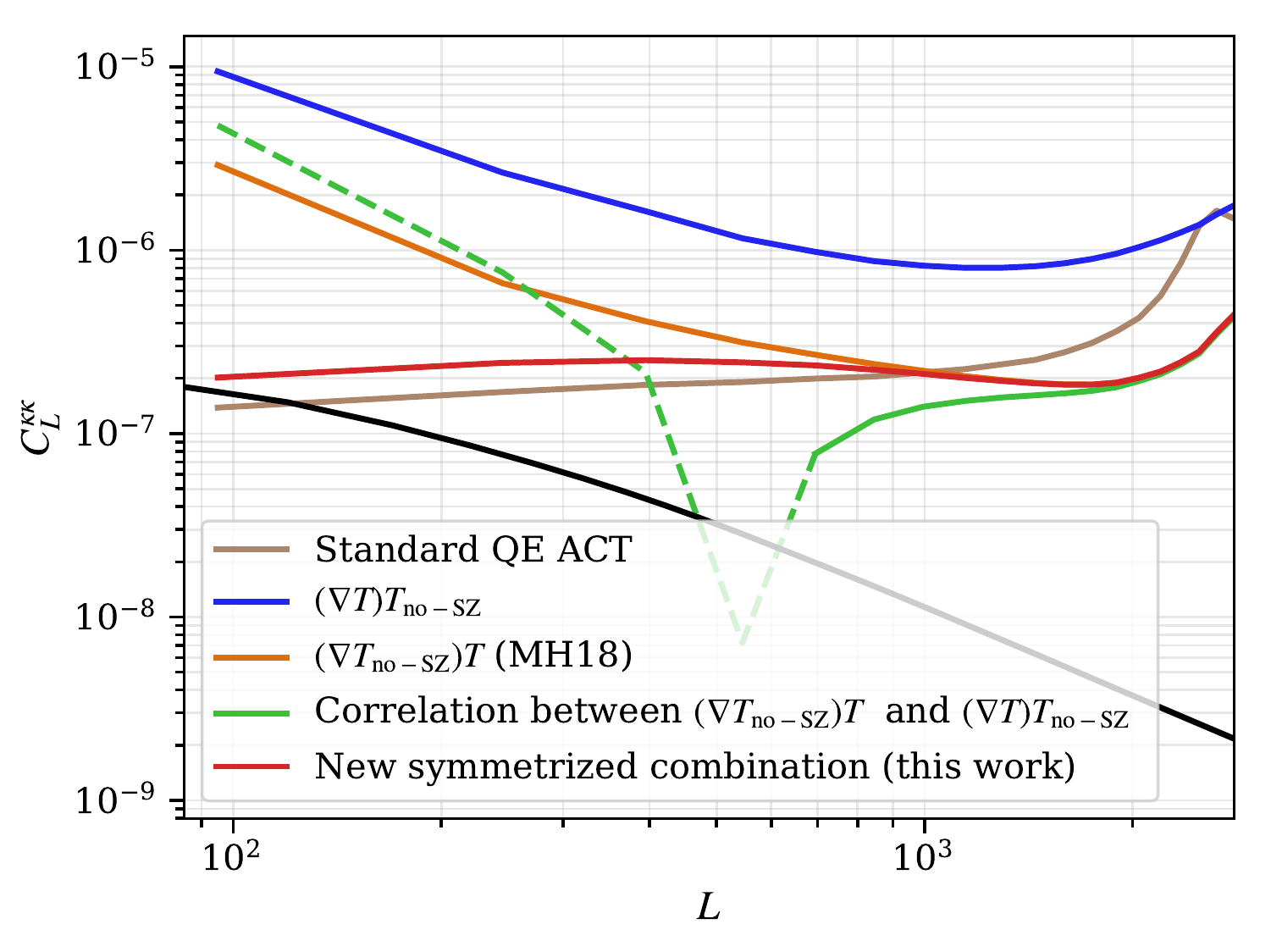}
    \caption{The noise power per mode for the temperature-only estimator for different cases in the {\tt D56} region. This plot shows how the
      symmetric cleaned estimator presented in this work lowers the noise
      compared to the asymmetric estimator. The green curve shows the
      cross-noise between the two different asymmetric estimators, with negative
      values in dashed. The anti-correlation of the noise on large scales
      between the two different asymmetric estimators leads to a cancellation in
      the optimal co-add of these that results in the red noise curve for our
      new symmetric cleaned estimator, which recovers the forecast performance in MH18.}
    \label{fig:noiseforecastsTT}
\end{figure}

The goal of this appendix is to illustrate the noise properties of the different
lensing estimators used in this work, with particular emphasis on the noise of
the new symmetric cleaned estimator that is free of tSZ contamination.

The estimated lensing convegence map in real space from a fixed polarization combination $XY$ for CMB maps is \citep[e.g.,][]{Hu:2007}:

\begin{equation}
    \hat\kappa^{XY}(\mathbf{\hat{n}})=\int \frac{d^2\Lb}{(2\pi)^2}e^{i\Lb\cdot \vec{n}}\hat\kappa^{XY}(\Lb)
\end{equation}
with

\begin{equation}
        \hat{\kappa}^{XY}(\Lb)=-A_L^{XY}\int d^2\mathbf{\hat{n}}e^{-i\mathbf{\hat{n}}\cdot\ellb}{\rm Re}\{\nabla\cdot [\vec{G}^{XY}(\mathbf{\hat{n}})L^{Y*}(\mathbf{\hat{n}})]\}
    \label{eq:kappaLreal}\ ,
\end{equation}
where  $XY \in \{ T_iT_j, T_iE_j, E_iE_j, E_iB_j \}+{i \leftrightarrow j}$ with the indices characterizing maps with different data content (e.g. from different experiments or with different component separation techniques),
$A_L^{XY}$ is a normalization to ensure that we recover an unbiased estimate of the convergence field, and  $\vec{G}^{XY}(\mathbf{\hat{n}})$ and $L^{Y*}(\mathbf{\hat{n}})$ are filtered versions of CMB maps. The details of these filtered maps can be found in \citet{Hu:2007}.

The normalization is
\begin{equation}
    A_L^{XY}=\frac{L^2}{2}\left[\int \frac{d^2\ellb}{(2\pi)^2}(\Lb\cdot \ellb) W_l^{XY}W_{|\Lb-\ellb|}^{Y}f_{XY}(\ellb, \Lb-\ellb)\right]^{-1} 
\end{equation}{}
where $W^{XY}_{l}, W^{X}, f_{XY}(\vec{l}, \vec{L}-\vec{l})$ can be found again in \citet{Hu:2007}. The lensing convergence estimator expands to

\begin{equation}
    \label{eq:convergence}
    \hat{\kappa}^{XY}(\Lb)=A_L^{XY}\int \frac{d^2\ellb}{(2\pi)^2}(\Lb\cdot \ellb) W_l^{XY}X(\ellb)W_{|\Lb-\ellb|}^{Y}Y(\Lb-\ellb) \ .
\end{equation}

The covariance of this estimator, $N^{XY,WZ}(\Lb)$ is

\begin{equation}
   \begin{split}
     \langle \hat{\kappa}^{XY}(\Lb)\hat{\kappa}^{WZ}(\Lb')\rangle_{CMB}- \langle \hat{\kappa}^{XY}(\Lb)\rangle_{CMB}\langle\hat{\kappa}^{WZ}(\Lb')\rangle_{CMB}= \\
     =(2\pi)^2\delta_D^{(2)}(\Lb-\Lb') A_L^{XY}A_L^{WZ*}\int \frac{d^2\ellb}{(2\pi)^2}(\Lb\cdot \ellb)W_l^{XY}W_{|\Lb-\ellb|}^Y\times \\
   \times [(\Lb\cdot \ellb)W_l^{WZ}W_{|\Lb-\ellb|}^ZC_l^{\Bar{X}\Bar{W}}C_{|\Lb-\ellb|}^{\Bar{Y}\Bar{Z}}+ \\
        +(\Lb\cdot (\Lb-\ellb))W_{|\Lb-\ellb|}^{WZ}W_{l}^ZC_l^{\Bar{X}\Bar{Z}}C_{|\Lb-\ellb|}^{\Bar{Y}\Bar{W}}]\ .
  \end{split}
\end{equation}

\begin{figure}
    \centering
    \includegraphics[width=9cm]{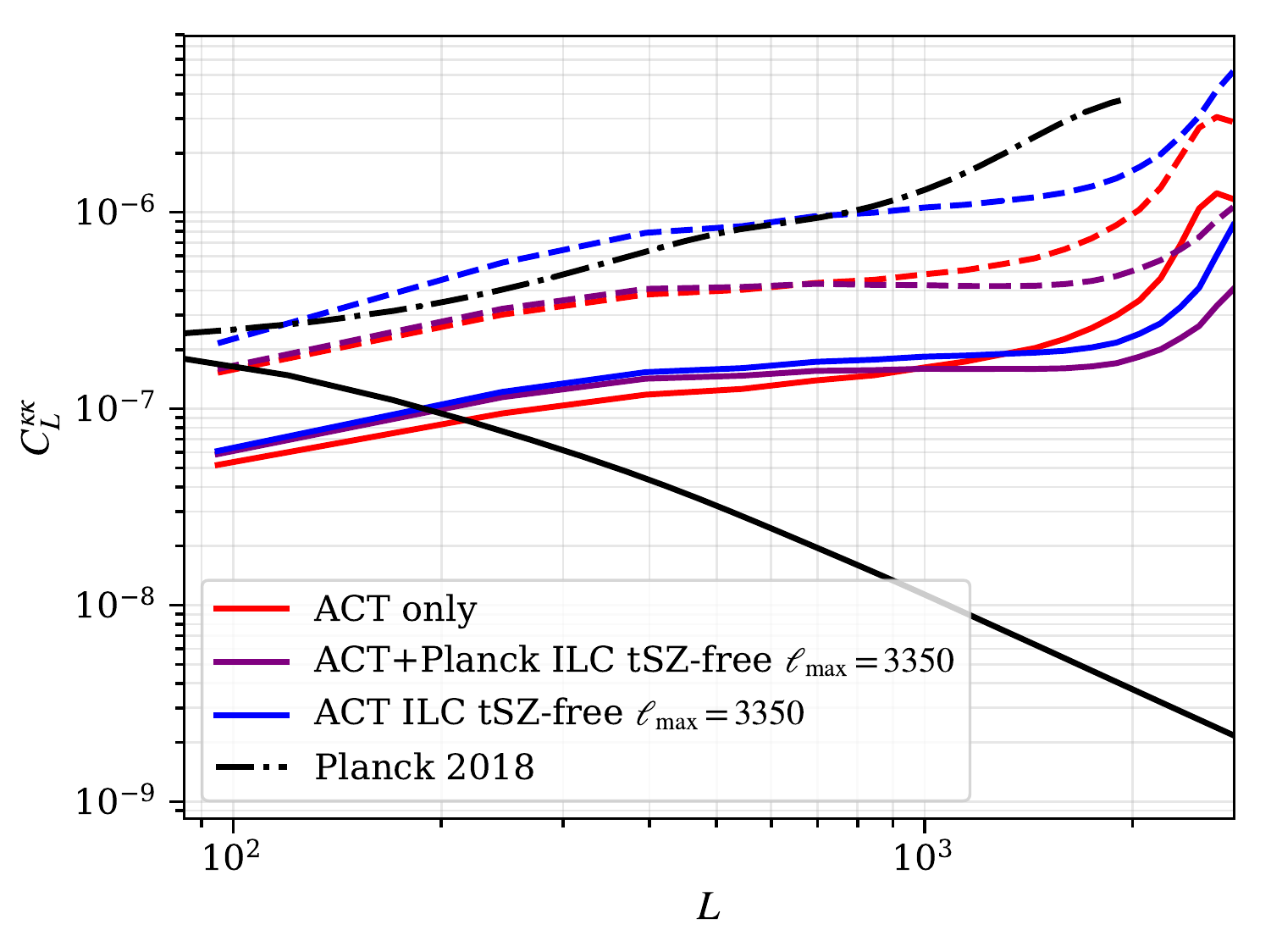}
    \caption{The noise power per mode in our maps for different patches for the minimum variance (temperature + polarization) co-add of the CMB lensing maps from this work. Solid colored lines represent {\tt D56}, dashed lines represent {\tt BN} and dashed and dotted represent \emph{Planck} 2018. The theory expectation for the signal is shown in black. Our maps are signal dominated for $L<100$ in {\tt BN} and $L<200$ in {\tt D56}.}
    \label{fig:noiseforecastsmv}
\end{figure}

When the maps involved are identical ($X=Y$, e.g. for TT and EE estimators where both fields have the same data), the minimum-variance filters have a simple form as shown in \cite{Hu:2007} and the estimator can be written in a separable manner (i.e., can be written using sums of products of a function of $\ellb_1$ times a function of $\ellb_2$) that allows for fast evaluation with FFTs. Moreover, the estimator variance ($X=Y=W=Z$ above) has a simple relation to the normalization $N_L \propto A_L L^2$. This no longer holds when $X \neq Y$. In particular, for our case of interest where we mix maps with different component separation techniques, $X=T_{\rm no-fg}$ and $Y=T_{\rm with-fg}$, the minimum variance estimator does not have a simple separable form. MH18 used an approximation to the minimum-variance estimator that consisted of the two different maps being independently Wiener filtered. When the weights in the estimator are not minimum-variance, the relation (assumed in the forecast of that paper) that $N_L \propto A_L L^2$ no longer holds. The true performance is the orange curve in Figure B1. However, a simple heuristic extension of the MH18 estimator recovers performance close to what was forecast there: the two asymmetric estimators  $\hat\kappa(T_{\rm no-fg}, T_{\rm with-fg})$, $\hat\kappa(T_{\rm with-fg}, T_{\rm no-fg})$ combined in a minimum variance combination $\hat\kappa^{TT}_{\rm symm,fgfree} = \sum w_\alpha(\Lb)\hat{\kappa}_{\alpha}(\Lb)$ with weights given by Eq. \ref{eq:weights}, where $\alpha \in \{(T_{\rm no-fg}T_{\rm with-fg}), (T_{\rm with-fg}T_{\rm no-fg})\}$ and $N^{-1}$ the inverse of the $2 \times 2$ covariance matrix taking into account the cross-correlation between the two estimators.

 In Figure \ref{fig:noiseforecastsTT} we show the noise curves for this $TT$
 symmetric cleaned estimator, as well as the asymmetric estimators. In Figure
 \ref{fig:noiseforecastsmv} we show lensing minimum variance noise curves, which
 include polarization lensing measurements. These are shown for three different
 cases that differ in how the TT estimator is calculated (a) using the tSZ-free
 symmetric cleaned estimator with both \emph{Planck} and ACT data combined with ILC
 (our baseline, in purple)  (b) using only ACT data with the $1/N$ co-adding
 scheme, and no deprojection of foregrounds (red) and (c) using the tSZ-free
 symmetric cleaned estimator with only ACT data combined with ILC (blue).

%-------------------------------------------------------------------
%\clearpage
\bibliographystyle{mnras} 
\bibliography{biblio}

\end{document}